\documentclass{aa}
\usepackage{graphicx}
\usepackage{txfonts}
\usepackage{natbib}
\bibpunct{(}{)}{;}{a}{}{,} 

\def\aaps{A\&AS}
\def\aap{A\&A}
\def\apj{ApJ}
\def\apjs{ApJS}
\def\aj{AJ}

\def \hi {\ion{H}{i}}

\def\kms{km\,s$^{-1}$}

\def\deg{\hbox{$^\circ$}}
\def\arcmin{\hbox{$^\prime$}}

\def\fdg{\hbox{$.\!\!^\circ$}}
\def\farcm{\hbox{$.\mkern-4mu^\prime$}}

\defcitealias{Naomi2009}{Paper~I}
\defcitealias{Kalberla2010}{Paper~II}

\begin{document}

   \title{GASS: The Parkes Galactic All-Sky Survey.}

   \subtitle{Update: improved correction for instrumental effects and new
     data release}

\author{P.\ M.\ W.\ Kalberla,\inst{1}
         and U. Haud \inst{2}}

\institute{Argelander-Institut f\"ur Astronomie, Universit\"at Bonn, Auf
  dem H\"ugel 71, 53121 Bonn, Germany,\\
  \email{pkalberla@astro.uni-bonn.de}
\and
   Tartu Observatory, 61602 T\~oravere, Tartumaa, Estonia,
   \email{urmas@aai.ee}
}

   \authorrunning{P.\,M.\,W. Kalberla \& U.Haud}

   \titlerunning{GASS, Third Data Release}

   \offprints{P.\,M.\,W. Kalberla}

   \date{Received 11 February 2015 / Accepted 24 April 2015}

  \abstract
  {The Galactic All-Sky Survey (GASS) is a survey of Galactic atomic
     hydrogen (\hi) emission in the southern sky observed with the
     Parkes 64-m Radio Telescope. The first data release (GASS I)
     concerned survey goals and observing techniques, the second release
     (GASS II) focused on stray radiation and instrumental corrections.}
  {We seek to remove the remaining instrumental effects and present a
     third data release.}
  {We use the HEALPix tessellation concept to grid the data on the
     sphere. Individual telescope records are compared with averages on
     the nearest grid position for significant deviations. All averages
     are also decomposed into Gaussian components with the aim of
     segregating unacceptable solutions. Improved priors are used for an
     iterative baseline fitting and cleaning. In the last step we
     generate 3-D FITS data cubes and examine them for remaining
     problems.}
  {We have removed weak, but systematic baseline offsets with an
     improved baseline fitting algorithm. We have unraveled correlator
     failures that cause time dependent oscillations; errors are mostly
     proportional to the observed line intensity and cause stripes in
     the scanning direction. The remaining problems from radio frequency
     interference (RFI) are spotted. Classifying the severeness of
     instrumental errors for each individual telescope record (dump)
     allows us to exclude bad data from averages. We derive parameters
     that allow us to discard dumps without compromising the noise of
     the resulting data products too much. All steps are reiterated
     several times: in each case, we check the Gaussian parameters for
     remaining problems and inspect 3-D FITS data cubes visually. We
     find that in total {$\sim1.5$}\% of the telescope dumps need to be
     discarded in addition to {$\sim0.5$}\% of the spectral channels
     that were excluded in GASS II.}
   {The new data release (GASS III) facilitates data products with
      improved quality. A new web interface, compatible with the
      previous version, is available for download of GASS III FITS cubes
      and spectra.}
  \maketitle
%

\section{Introduction}
\label{Intro}

The Parkes Galactic All-Sky Survey (GASS) maps the Galactic atomic
hydrogen emission in the southern sky for declinations $\delta \la 1
\degr$. The observations were made between January 2005 and October 2006
with the multibeam system on the 64-m Parkes Radio Telescope.
\citet[][Paper I]{Naomi2009} described the survey goals and observing
techniques in detail. All of the initial data reduction and imaging for
the first data release (GASS I) is also described in
\citetalias{Naomi2009}.

Observations of the Galactic \hi~ emission are affected by stray
radiation received by the antenna diagram, and a correction is
mandatory. The second data release (GASS II) was published after proper
corrections became available \citep[][Paper II]{Kalberla2010}. 
At this stage it also turned out that part of the data
was severely affected by radio frequency interference (RFI). About 0.5\%
of the data were discarded \citep[see also][]{Kalberla2011} , but some
RFI remained.

The GASS II RFI mitigation algorithm was based on median filtering,
searching for sharp gradients in each individual observed spectrum.
Spikes are typical for RFI but strong gradients can also have natural
origin, such as absorption components and unresolved small scale
fluctuations in the \hi~ distribution. This small scale structure can
exist along filaments, and tests during the preparation of GASS II have
shown that the applied median filtering could already degrade genuine
emission features at low brightness temperatures $T_B \sim2 $ K. The RFI
excision was therefore mostly restricted to channels with $T_B \la 0.5 $
K, a very conservative approach with the aim of not impacting any
genuine \hi~ emission features. Despite some remaining problems,
priority was given to rapid publication.

Working with data products from GASS II revealed problems that made it
desirable to improve the data processing in several respects:
\begin{enumerate}
   \item Data may be affected by the remaining RFI, mostly in emission
      line regions with $T_B \ga 0.5 $ K.
   \item Maps can be degraded by stripes in the scanning direction,
      either right ascension or declination, predominantly in regions
      with strong \hi~ emission.
   \item Baselines show occasional faint but systematic offsets, mostly
      along the borderline of the Galactic disk but also for isolated
      faint emission line features at high velocities.
   \item Emission lines and instrumental baselines at positions with
      strong continuum emission are partly unreliable.
   \item Some emission from local galaxies is badly treated by the
      baseline algorithm.
\end{enumerate}
We decided to address these issues and, in the following, we present our
investigations that led to a new data release (GASS III).

\section{The iterative correction algorithm}

\subsection{General strategy}
\label{General}

Dealing with instrumental problems is often a complex and difficult
task. Individual strategies may fail to remove all problems in a single
step. It is then necessary to iteratively repeat the approaches, step by
step with an improved algorithm. This kind of strategy is, in case of
the GASS, enforced by the fact that we are dealing with $2.8 \,10^7$
spectral dumps for individual receivers and polarizations, each with 5 s
integration time. This number does not allow for a visual inspection or
correction of all individual dumps. Yet, the corrected data product
needs to be inspected for remaining failures \citep[The proof of the
pudding is in eating,][]{Cervantes}. We have done this in two ways,
checking average \hi~ profiles and 3-D FITS cubes of the whole sky.

For a correction of instrumental problems our basic assumption is that
only a fraction of the data is affected. The survey had two independent
coverages, and therefore instrumental defects should mostly be spatially
uncorrelated. Averaging these coverages should then diminish
instrumental problems, in particular, if problematic data can be
excluded from the average accordingly. Thus the first step of each
iteration cycle is the generation of average profiles; our initial
estimate was derived from GASS II. These averaged data can also be
searched for remaining problems. We decompose all profiles into Gaussian
components and check the results for outliers that might indicate
instrumental problems.

To generate a new dump database, we use calibrated telescope data that
have been corrected for stray radiation. These data are then corrected
for instrumental baselines, as described in \citetalias{Kalberla2010},
except that we now use the previously determined average profiles to
derive an initial estimates of the baseline fit. The data are then
cleaned for RFI as described in \citetalias{Kalberla2010}. Each
individual dump is compared with the nearest profile from the previously
determined average, the rms deviation is recorded as a measure of the
quality of the dump. Excluding bad data, we eventually generate 3-D FITS
cubes for visual inspection, and we also proceed to calculate a new
version of improved averages.

It is obvious that these iterations can only be successful if most of
the data are of good quality and if other reliable criteria to define
data as ``good'' or ``bad'' can be included without affecting features
of the \hi~ brightness temperature distribution.

\subsection{The HEALPix database}
\label{HEALPix}

For a database containing average emission profiles, we would need to
define a grid with constant distances to neighboring positions all over
the sky. This scheme does not exist, but the Hierarchical Equal Area
isoLatitude Pixelization \citep[HEALPix, ][]{Gorski2005} is a close
approximation. HEALPix is a versatile structure for the pixelization of
data on the sphere that also enables an easy comparison for data with
different resolution. A single parameter, $ N_\mathrm{side}$, defines
the pixel resolution $ \theta = \sqrt{3/\pi} \ 3600\arcmin /
N_\mathrm{side} $ as well as the total number of pixels $N_\mathrm{pix}
= 12 N_\mathrm{side}^2$ on the sphere. We have chosen $ N_\mathrm{side}
= 1024 $, corresponding to a resolution of $ \theta = 3\farcm44, $ which
is small in comparison to the average full width at half maximum (FWHM)
GASS beamwidth of $14\farcm4$. From the flagged dump database, we
calculate all profiles on the HEALPix grid in each step of the
iterations. Here we also include data available in the overlap regions
for declinations $ \delta \ga 1 \degr $. For the GASS, the HEALPix
database contains 6\,655\,155 profiles, 6\,401\,256 of these at $\delta
\le 1 \degr$.

To generate the HEALPix database, we use a Gaussian convolution with a
FWHM width of $7\farcm5$ as the kernel. This results in an effective
resolution of FWHM $16\farcm2$ for the average spectra. The HEALPix
database is updated every iteration cycle, excluding data that are
affected by instrumental problems (see Sections \ref{Clean} to
\ref{Flagging}). In the first step, we used the GASS II database.

When calculating the weighted arithmetic mean profile $T_B(v)$ on the
HEALPix grid from $N$ available dumps, we take into account for each
channel $j$ at the velocity $v_j$ that data may need to be excluded
because of previous flagging, setting either $f_{i,j} = 0 $ or $f_{i,j}
= 1$. Here the index $i$ counts the observed dumps that are used for the
average,

\begin{equation}
T_B(v_j) = \frac{\sum_{i=1}^N w_i f_{i,j} T_i(v_j)}{\sum_{i=1}^N
  w_if_{i,j}}
 \label{Eq1}
.\end{equation}
Here $w_i$ is the weight from the Gaussian kernel function. To trace the
quality of individual averaged brightness temperature profiles we record
the weights,
\begin{equation}
W(v_j) = \sum_{i=1}^N w_if_{i,j}
 \label{Eq2}
,\end{equation}
and the rms scatter $S(v_j)$ from the unbiased weighted estimator
\begin{equation}
    S^2(v_j) = \frac{\sum_{i=1}^N w_i f_{i,j} T_i^2(v_j) \cdot \sum_{i=1}^N
  w_i f_{i,j} - (\sum_{i=1}^N w_i f_{i,j} T_i(v_j))^2}{(\sum_{i=1}^N w_i f_{i,j})^2
  - \sum_{i=1}^N (w_i f_{i,j})^2}.
 \label{Eq3}
\end{equation}

\subsection{Gaussian decomposition}
\label{Gauss}

After generating the HEALPix database, we decompose all brightness
temperature spectra $T_\mathrm{B}$ into Gaussian components. The aim of
the decomposition is quality control through detecting outliers in the
distributions of the Gaussian parameters. These outliers may indicate
instrumental problems. For the decomposition we use mostly the same
approach, which was described by \citet{Hau00} and applied earlier to
the Leiden/Argentine/Bonn (LAB) data \citep{Kalberla2005}. In general,
this is a rather classical Gaussian decomposition, but with two
important additions. First of all, our decomposition algorithm does not
treat each \ion{H}{i} profile independently, but assumes that every
observed profile shares some similarities with those observed in the
neighboring sky positions. In addition, besides adding components into
the decomposition, our algorithm also analyzes the results to find the
possibilities for removing or merging some Gaussians without reducing
too much the accuracy of the representation of the original profile with
decomposition. We utilized both of these means for reducing the
ambiguities inherent to the decomposition procedure.

For the decomposition of the GASS, we also introduced some modifications
to our old algorithm. First of all, the accuracy of the measured
brightness temperatures is not the same for all profile channels. The
noise level depends on the signal strength and different corrections
introduce additional uncertainties. During the decomposition all this is
considered through the weights $W_\mathrm{GD}(v_j)$, assigned to all
values of $T_\mathrm{B}(v_j)$. In the LAB survey, there was in general
only one observed \ion{H}{i} profile per sky position and we had to
estimate $W_\mathrm{GD}(v_j)$ mostly theoretically \citep[Eq. 2
in][]{Hau00}. For every HEALPix pixel of the GASS, the profile is found
as an average of some tens of original spectral dumps (typically $N$ =
40 to 60). As a result, for each individual channel $j$, besides the
weighted mean brightness temperature $T_\mathrm{B}(v_j)$ (Eq. \ref{Eq1})
we also have the rms deviation $S(v_j)$ (Eq. \ref{Eq3}) from the average
and the weight $W(v_j)$ (Eq. \ref{Eq2}). We can use this information on
the statistical reliability of the average values in each profile
channel and calculate for the decomposition the weights of the channels as,

\begin{equation}
   W_\mathrm{GD}(v_j)=W(v_j)/S^2(v_j).
   \label{Eq4}
\end{equation}

Another change in the decomposition algorithm is made available by the
considerably increased power of the computers. In the LAB survey,  we
used the decomposition results of one of the neighboring profiles of any
given profile as an initial estimate of the Gaussian parameters, and we
only started the decomposition of the first profile with one roughly
estimated component at the highest maximum of the profile. In the GASS
study, the decomposition is divided into two stages. In the first run,
we decompose all profiles in the HEALPix database independent of its
neighbors, starting with one Gaussian at the brightest tip of the
profile. In the second stage, we compare the results for neighboring
profiles. To accomplish this, we use the decomposition, obtained so far
for each profile, as an initial approximation for all eight nearest
neighbors of this profile, and check whether this leads to better
decomposition of these neighboring profiles. If the decomposition of a
neighbor is improved, the new result is used as an initial solution for
the eight neighbors of this profile, and so on. The process is repeated
until no more improvements are found (of the order of 500 runs through
the full database).

During this process, we estimate the goodness of the obtained fit using
two criteria. First of all, for the acceptable decomposition, the rms of
the weighted deviations of the Gaussian model from the observed profile
must be no more than 1.004 times the weighted noise level of the
emission-free baseline regions of this profile. The value of the
multiplier has been chosen so that the averages of the noise levels and
the rms deviations of the models over all HEALPix profiles of the survey
are equal (in practice, for the final decomposition the difference of
these averages is less than 0.001\% of their mean). If for some profiles
the rms criterion is satisfied for more than one trial decomposition
with different numbers of Gaussians, we accept the solution with the
smallest number of the components. In the case of acceptable
decompositions with equal numbers of the Gaussians, we chose the
decomposition with the smallest rms as the best. The decomposition
process is described in more detail in Secs. 3.1. and 3.2. of
\citet{Hau00}.

Because of the described usage of the neighboring profiles, it is also
important to decompose the profiles at $\delta > 1\degr$, but the
corresponding results are not usable for the following analysis. On the
edge of the observed area at $\delta \approx 3\degr$, each profile has
fewer neighbors than inside the field. For these border profiles, the
program has to use a smaller number of different initial approximations
for the decomposition. In addition, since the scan pattern of the
multibeam system covers $\Delta\delta \sim 2\degr$ (see Fig. 1 of
\citetalias{Naomi2009}), the sky coverage becomes incomplete for $1\degr
\la \delta \la 3\degr$ and the total weights for the averaged profiles
decrease for increasing declination. Both these circumstances reduce the
quality of the decomposition results near the edge of the observed
region of the sky. Therefore, after decomposing 6\,655\,155 profiles, we
analyze for quality only the 6\,401\,256 profiles at $\delta \le 1
\degr$. These averaged profiles already have the complete sky coverage
and are decomposed with the same thoroughness.

Finally, in the case of the LAB, we only used positive Gaussians (the
central height of all components $T_\mathrm{BC} > 0~\mathrm{K}$) for
decomposition. The parts of the profiles with $T_\mathrm{B} <
0~\mathrm{K}$ were not considered at all. With the GASS, we also fit
negative Gaussians to the regions of the profiles, where the brightness
temperature is on average below zero. These regions with $T_\mathrm{B} <
0~\mathrm{K}$ may be caused mainly by two different phenomena: the real
\ion{H}{i} absorption or the incorrectly determined baseline. For
quality control of the profiles, only the Gaussians caused by the
baseline problems are important. We try to distinguish them from the
negative components, representing the absorption. Concerning the latter
features, we stress that for fitting the regions of the profiles, where
$T_\mathrm{B} \ge 0~\mathrm{K}$, we still only use positive Gaussians
and not a combination of positive and negative components. Therefore,
only very strong absorption, where the $T_\mathrm{B}$ of the profile
drops below the continuum level, induce negative Gaussians. In
general, absorption is modeled as a cap between positive Gaussians.

The described decomposition of the full HEALPix database of the GASS
profiles requires about 3.2 days on 8 cores of Dell R910 server. When we
started on the first iteration after GASS II, the decomposition gave us
on average 9.17 Gaussians per profile. For the final data of GASS III,
this number was 9.08. The decrease is achieved on account of the
reduction of the spurious features in the profiles. Actually, the better
indicator of the improvement is the fact that the program was allowed to
use up to 50 Gaussians per profile during decomposition. This number of
components was used only in rare cases of obviously bad profiles or for
some very complicated profiles near the Galactic center. This happened
662 times in the initial and in 19 cases in the final data set.

\subsection{Baseline fitting}
\label{Baseline}

In \citetalias{Kalberla2010}, the LAB survey data \citep{Kalberla2005}
were taken to derive initial parameters for the baseline fit. For this
purpose the nearest LAB profile was smoothed to an effective resolution
of 5 \kms~ and channels where $T_\mathrm{LAB} < 0.9$K were used as a
first guess of emission-free baseline regions. This prior works well for
diffuse \hi~ emission regions, but can fail seriously in presence of
small scale features. An example is baseline fitting at positions with
small \hi~ clumps or emission from other galaxies. For GASS II some
provision was made during baseline fitting to recover small scale
structure, but success was limited by the fact that individual dumps
with 5s integration time are noisy, typical $\sigma_\mathrm{rms} \sim
0.4$ K. The algorithm only allows us to recover small scale structure at
a $3 \sigma$ detection level. To overcome these limitations of the GASS
II data reduction, it is necessary to use more sensitive data at full
spatial resolution.

To generate a clean baseline for individual dumps of the GASS III
database, we use the nearest profile from the HEALPix database. Channels
with $T_\mathrm{GASS} < 3 \sigma_\mathrm{rms}$ are used as a first guess
of emission-free baseline regions. Flags, indicating RFI or other
defects, are considered. The fits improved significantly in comparison
to the GASS II version, but we still found notable problems for baseline
fitting in the presence of external galaxies. At these positions, we
needed to include parameters from the HIPASS Bright Galaxies Catalog
\citep{Koribalski2004}. We repeat iteratively the baseline fitting for
all survey data with the changes described above, starting with the same
stray-radiation corrected database as in \citetalias{Kalberla2010}.

\subsection{Flagging individual channels for RFI}
\label{Clean}

For flagging and cleaning of channels that were affected by RFI, we use
the same methods as described in Sect. 5 of \citetalias{Kalberla2010}.
In brief, we apply median filtering to detect and flag RFI spikes. To
prevent generic \hi~ features to be affected by this filtering process,
we limited the search algorithm to brightness temperatures $T_B \la 0.5
$ K. Also dumps close to strong continuum sources were excluded within a
radius of 0.5\degr, see Sect. \ref{Flagging} for details. Flags
$f_{i,j}$ are recorded in the dump database.

\subsection{Flagging bandpass ghosts}
\label{ghosts}

To maximize sensitivity and recover all the extended \hi~ emission, the
GASS was observed by using in-band frequency switching with an offset of
3.125 MHz, corresponding to 660 \kms. The {\it Livedata} bandpass
correction (see \citetalias{Naomi2009}, Sect. 2.3.1) causes for every 
real emission line feature an associated,
spurious, negative image (``ghost'') in the other band, displaced by
$\pm 660$ \kms. These bandpass ghosts can be minimized by flagging all
channels that are affected by ghost features ($f_{i,j} = 0$). When
averaging profiles or making maps for these channels only half of the
data are available. This leads to an increase of the noise, but at least
systematic biases can be avoided this way.

\subsection{Correlator failures}
\label{Correlator}

A part of the GASS II data, predominantly at high brightness
temperatures, shows stripes in the scanning direction. The reason for
this failure remained a mystery for quite a while. Eventually, from the
Gaussian decomposition we detected that these problems were correlated
with channel-to-channel fluctuations that could be traced to individual
scans in the dump database. These channel-to-channel fluctuations tend
to increase with brightness temperature. These problems are time
dependent, hence, they show up as scanning stripes in a part of the data
(see Figs. \ref{Fig_RFI_A}, \ref{Fig_RFI_B}, and \ref{UFig04}).

Searching for a measure of these failures, we eventually found a very
simple approach. We independently add the intensities for all even and
odd channels of a dump. Typical differences for these sums are found to
be around 10 K\,km\,s$^{-1}$ and considerably higher values are
characteristic of correlator failures. The RFI causes strong
oscillations as well (see Fig. 5 of \citetalias{Kalberla2010}). Many of
these dumps, therefore, also get marked at this step.

\subsection{Excluding complete dumps}
\label{Flagging}

Flagging for RFI, as used for GASS II, was aimed to exclude individual
bad spectral channels. This affects about 0.5\% of all data. Here we
extended the algorithm to excise complete dumps.

While fitting instrumental baselines, we simultaneously determine rms
fluctuations $\sigma_\mathrm{rms}$ within the baseline region and store
these values in the profile header for later use. At the same time, we
record rms deviations between each individual dump and the nearest
average HEALPix profile. In both cases, the expected rms fluctuation
according to the integration time of 5s can be calculated. We record
these values in the profile header.

When calculating FITS maps and averaging profiles on the HEALPix grid,
we also exclude complete dumps whenever the rms noise
$\sigma_\mathrm{rms}$ in the baseline area exceeds the expected value by
a factor of 3 or alternatively, if the noise is low by a factor
  1/3. In terms of system noise,
this would imply unrealistic changes of the receiver temperature by
factors of 9. Dumps were also rejected if the rms difference between the
dump and the nearest HEALPix profile exceeded a factor of two of the
expected value. These differences do not necessarily imply RFI but can
also be caused by correlator failures, described in the previous
Section. To exclude these problems, we selected an acceptable limit of
50 K\,km\,s$^{-1}$ for the sum of the channel-to-channel fluctuations
and discarded all dumps above this $5 \sigma$ limit. For all profiles
with absorption lines, this limit was increased to 90 K\,km\,s$^{-1}$.

The factors applied as described above were tuned with the aim to
exclude only data with severe problems. Close to the Galactic plane we
noticed that these criteria led to an unacceptable increase of the noise
in the data products. For latitudes $|b| < 3\degr$ the acceptance level
for an increase of the noise within the baseline region was therefore
increased by a factor of two. For acceptable rms differences between
dumps and HEALPix profiles, we raised the clip level by 50\%.

For all dumps within a radius 0.5\deg around strong continuum sources,
the excision according to expected rms fluctuation was disabled since
such criteria are not applicable to absorption features. These sources
include those listed in Table 3 of \citet{Calabretta2014} in addition
RCW74, RCW145, and RCW146 from Table 2 of \citet{Rodgers1960} and HIPASS
J1324-42 from the HIPASS Bright Galaxies Catalog \citep{Koribalski2004}.

\section{Data Quality}
\label{Results}

During each iteration we checked our results for improvements. As
mentioned in Sects. \ref{Flagging} and \ref{Correlator}, bad dumps were
kept in the database but we stored quality indicators in the headers. It
was possible to tune the chosen criteria for best possible results when
creating the HEALPix database and when calculating 3-D FITS maps. In
both cases identical criteria are used for the rejection of bad data.

The final selection of clip levels for the exclusion of bad dumps was a
compromise between discarding affected dumps and retaining low noise in
the data products. Our first priority was to prevent any degradation of
generic \hi~ features. In the following, we demonstrate the use of
Gaussian decomposition and 3-D imaging for quality control.

\subsection{HEALPix profile badness}
\label{Bad}

\begin{figure}[!t]
   \centering
   \includegraphics[width=9.cm]{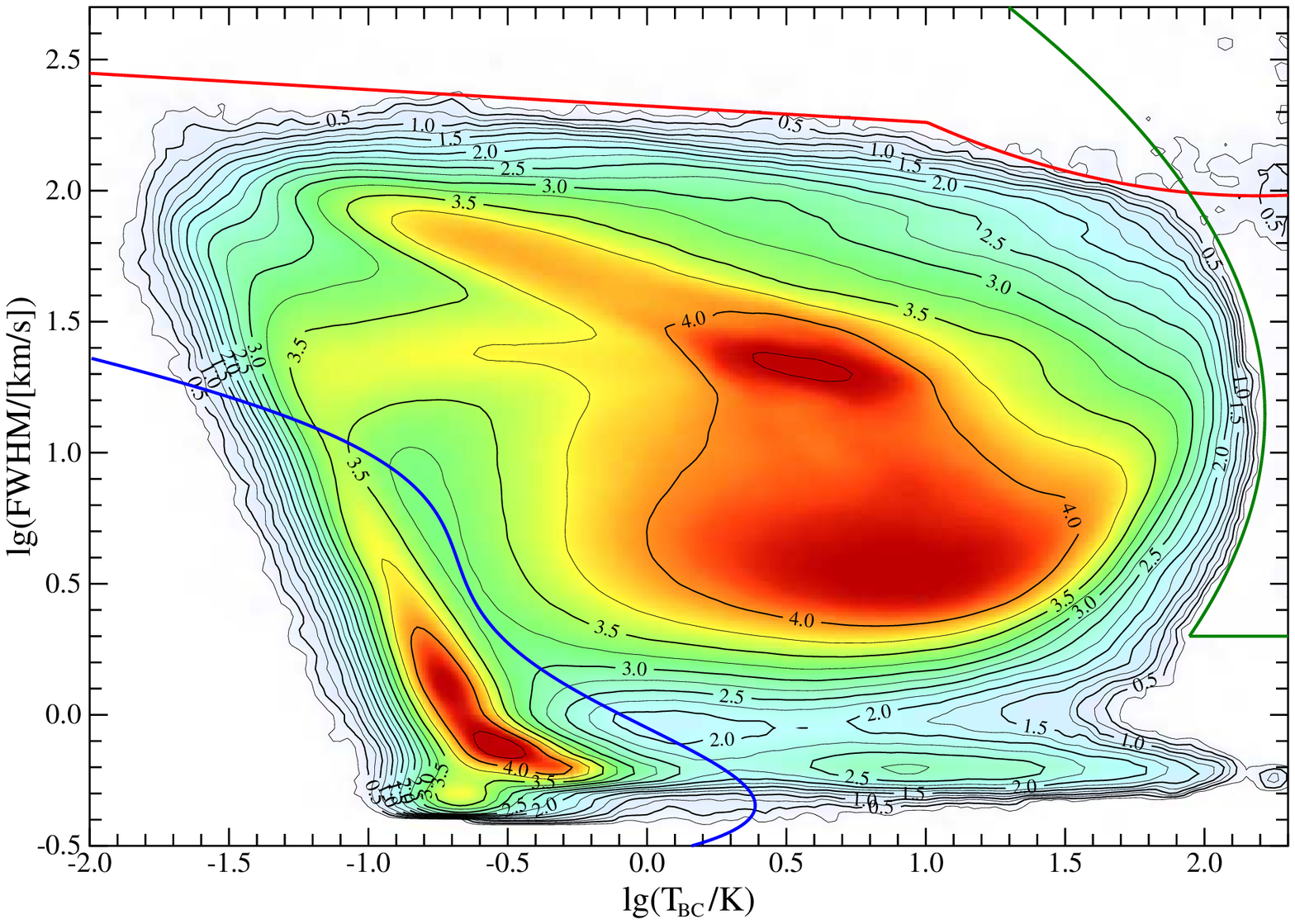}
   \includegraphics[width=9.cm]{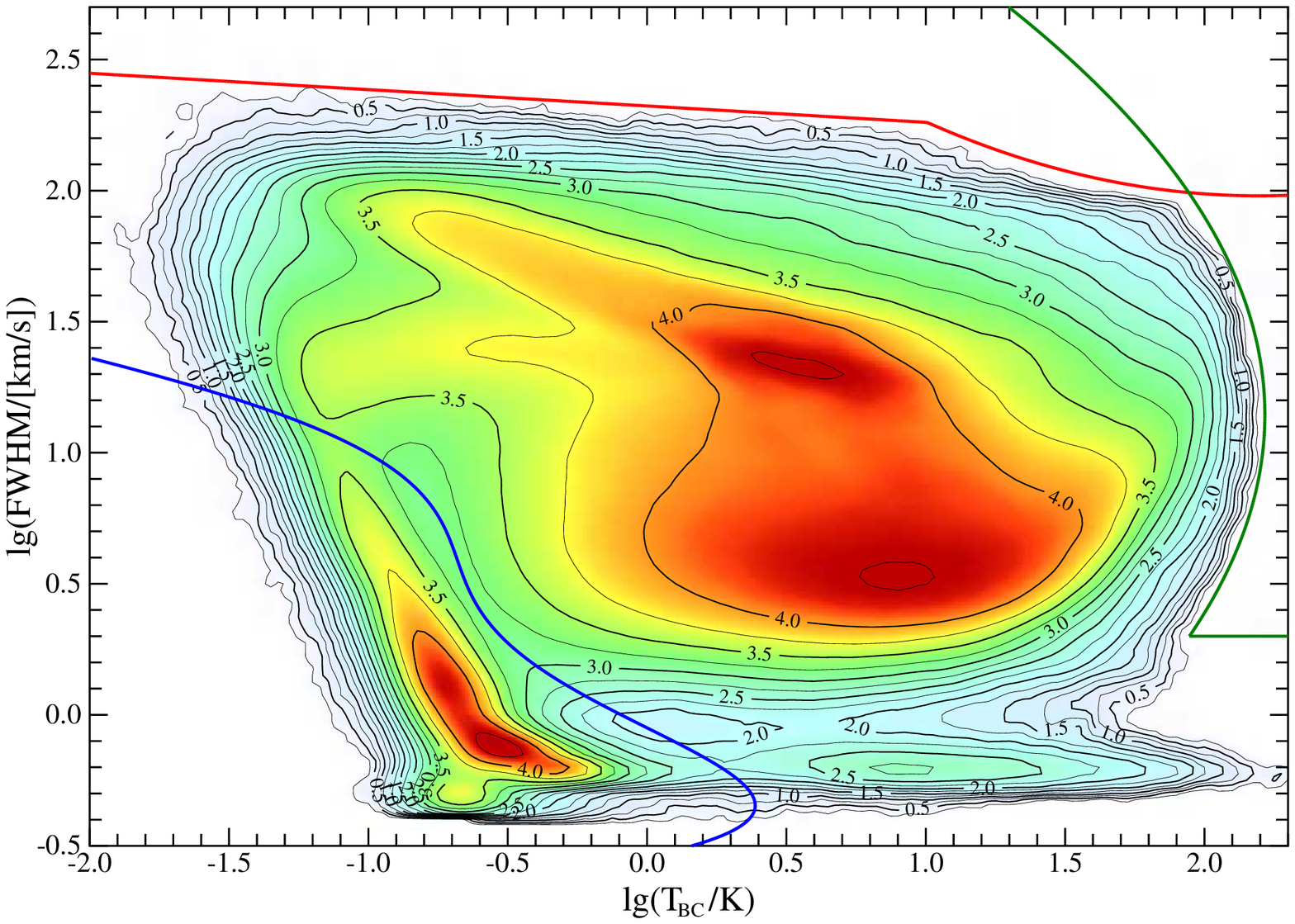}
   \caption{The distribution of the Gaussians in the
      $(\lg(T_\mathrm{BC}), \lg(\mathrm{FWHM}))$ plane. The upper panel
      is for the first and the lower is for the final iteration of the
      profile improvement. Isodensity lines are drawn in the scale of
      $\lg(N+1)$ with the interval of 0.25. The thick lines indicate the
      selection criteria used in the final iteration.}
   \label{UFig01}
\end{figure}

We use the results of the Gaussian decomposition to search for spurious
features in the averaged profiles of the HEALPix grid. To accomplish
this, we start with the frequency distributions of the obtained
parameters of the positive Gaussians in $(\lg(T_\mathrm{BC}),
\lg(\mathrm{FWHM}))$, $(V_\mathrm{C}, \lg(\mathrm{FWHM}))$ and
$(V_\mathrm{C}, \lg(T_\mathrm{BC}))$ planes (Figs.~\ref{UFig01} --
\ref{UFig03}). In the plots, we identify the regions, populated by the
outliers, which in one or another way differ from the components of the
main distribution and choose numerical criteria for the recognition of
the corresponding Gaussians. The definition of the selection criteria is
mainly based on the $(\lg(T_\mathrm{BC}), \lg(\mathrm{FWHM}))$ and
$(V_\mathrm{C}, \lg(\mathrm{FWHM}))$ plots, and the $(V_\mathrm{C},
\lg(T_\mathrm{BC}))$ plot is used mostly for additional checking. The
obtained criteria, together with those described in Sections \ref{Baseline},
\ref{Correlator} and \ref{Flagging}, were modified in every iteration.
Figs.~\ref{UFig01} -- \ref{UFig03} give the frequency distributions of
the Gaussian parameters in the initial (GASS II, upper panels) and final
(GASS III, lower panels) versions of the data together with the final
selection criteria (thick lines). We can see which kinds of Gaussians
are considered spurious. As illustrated by the lower panels of these
figures, in the final data the number of these Gaussians is reduced.

We used the parameters of the selected suspicious Gaussians to define a
profile badness with the aim of minimizing this parameter. The badness
is defined as a sum of the areas under these Gaussians. If a given
Gaussian is selected by more than one criterion, its area is considered
only once. The areas of the Gaussians are calculated for the velocities
covered by the observed profile. Of course, these criteria are rather
subjective, but they are used only for identifying the problematic
profiles and for the general assessment of the degree of contamination
of the averaged profiles with the spurious features. The modification of
the database is done on the basis of the study of the actual profiles,
as described in Sect. \ref{Maps}. The results from the Gaussian
decomposition are used as guides, pointing to the areas to study.
Therefore, the precise values of the criteria only influence the type
and number of the problems found, but do not directly influence the
corrections made in the database.

For the badness estimates of the profiles, we use six types of
criteria:

\begin{figure}[!t]
   \centering
   \includegraphics[width=9.cm]{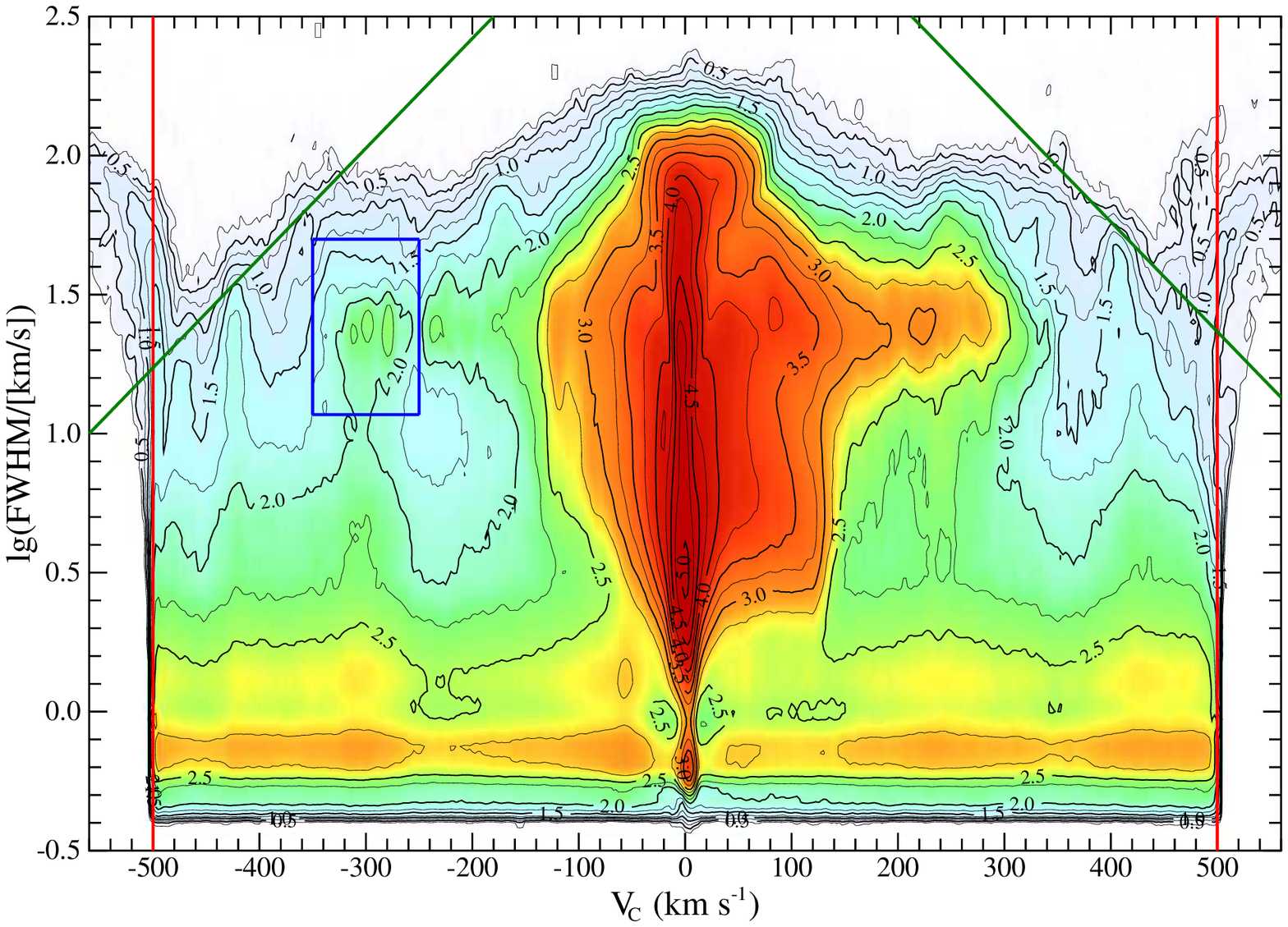}
   \includegraphics[width=9.cm]{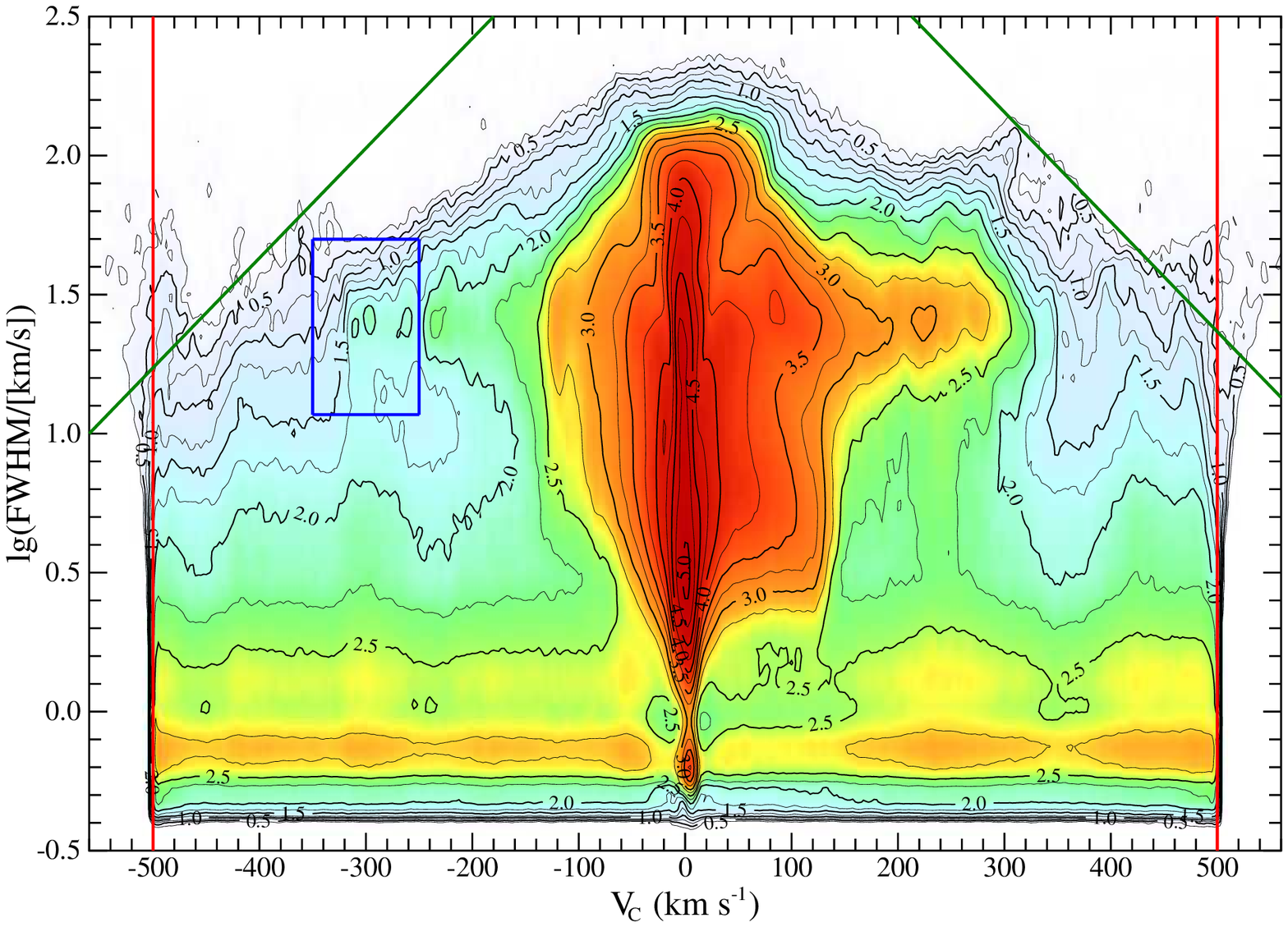}
   \caption{As Fig.~\ref{UFig01}, with the exception of the
      distribution of the Gaussians in the $(V_\mathrm{C},
      \lg(\mathrm{FWHM}))$ plane.}
   \label{UFig02}
\end{figure}

\begin{figure}[!ht]
   \centering
   \includegraphics[width=9.cm]{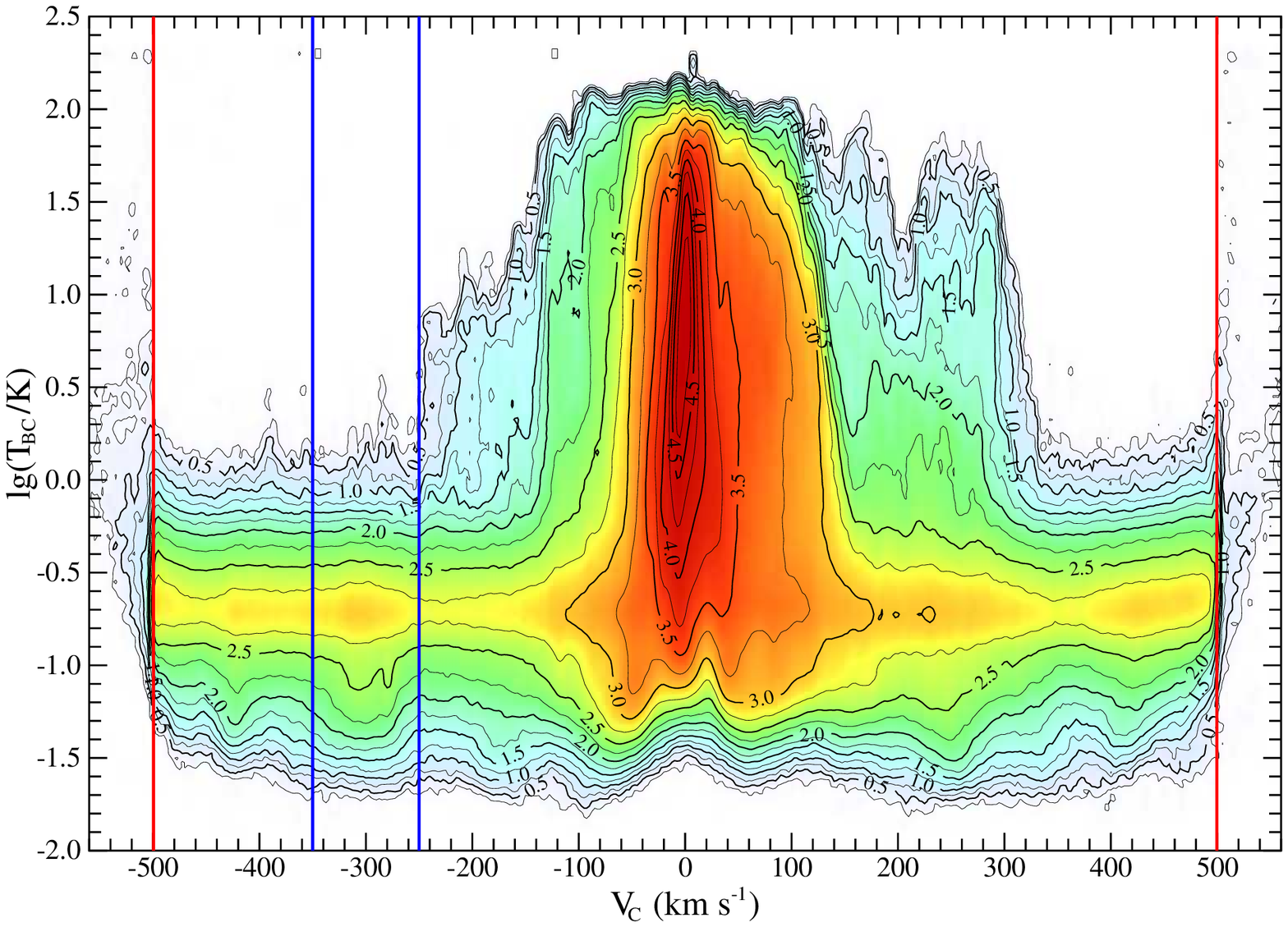}
   \includegraphics[width=9.cm]{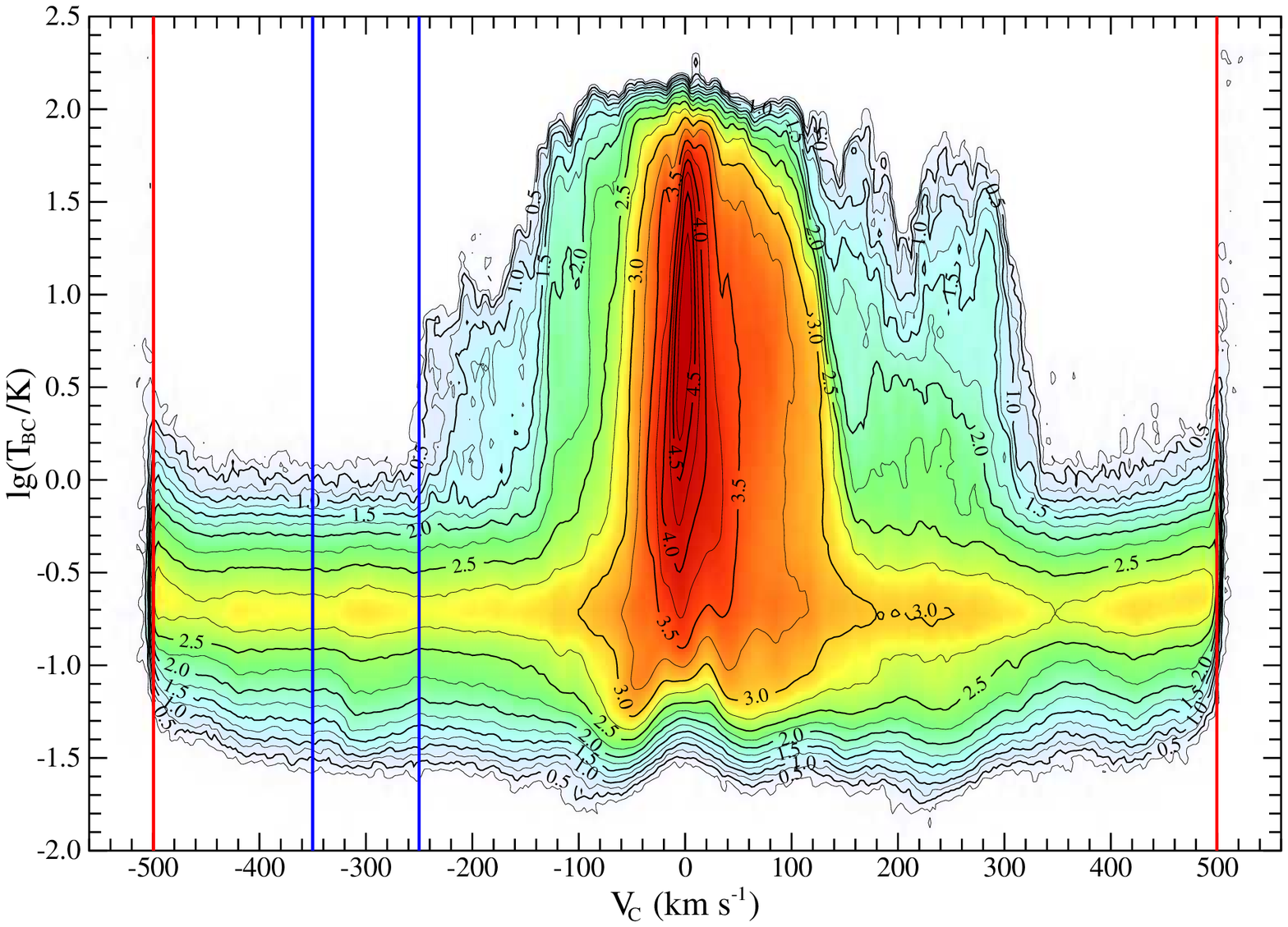}
   \caption{As Fig.~\ref{UFig01}, with the exception of the distribution
      of the Gaussians in the $(V_\mathrm{C}, \lg(T_\mathrm{BC}))$ plane.}
   \label{UFig03}
\end{figure}

\renewcommand{\labelenumi}{\roman{enumi}}
\begin{enumerate}

   \item This criterion is based on the negative Gaussians in the
      decomposition of the \ion{H}{i} profiles. It detects deviations of
      the profiles below the $T_\mathrm{B} = 0~\mathrm{K}$ line. Often
      these deviations are caused by an incorrect baseline. To protect
      real self-absorption features in the profiles near the Galactic
      center against rejection, we consider only the negative Gaussians
      with central velocities $|V_\mathrm{C}| > 5~\mathrm{km\,s}^{-1}$.
      However, sometimes other absorption features can be found. Most of
      the Gaussians, describing this absorption have $-70 < V_\mathrm{C}
      < 40~\mathrm{km\,s}^{-1}$, $T_\mathrm{BC} < -1~\mathrm{K}$ and
      $\mathrm{FWHM} < 12~\mathrm{km\,s}^{-1}$. We do not consider these
      components as spurious either. The areas under all other negative
      Gaussians are added to the badness of the corresponding profiles.

   \item The Gaussians, whose centers lie outside the velocity
      range covered by GASS, or whose centers are inside this range, but
      at the edge of the profile the Gaussian still has $T_\mathrm{B} >
      0.5~\mathrm{K}$. These Gaussians are mostly caused by a bad
      baseline near the bandpass edges. The in-band frequency switching
      technique, which causes significant uncertainties at high
      velocities, is responsible. The selection criterion is indicated
      in Figs.~\ref{UFig02} and \ref{UFig03} with the thick red vertical
      lines at $|V| = 500~\mathrm{km\,s}^{-1}$.

   \item The Gaussians above the slanted green lines in
      Fig.~\ref{UFig02}. Broad high-velocity profiles are not likely to
      be genuine since we know from HIPASS observations that this
      population of high-velocity clouds does not exist
      \citep{Putman2002}. Inspection of the profiles with such
      components has demonstrated that these Gaussians are frequently
      caused by baseline uncertainties near the profile edges and
      therefore this criterion somewhat extends, but partly duplicates
      our second criterion.

   \item The Gaussians above the red line in Fig.~\ref{UFig01} are
      mostly caused by two kinds of baseline problems. Some of them fit
      very wide but weak wings of the main emission peaks in the
      profiles, which may be caused by the incorrectly determined
      baseline near $V_\mathrm{C} = 0~\mathrm{km\,s}^{-1}$. This
      interpretation is supported by the fact that these wings are
      considerably reduced in the final data set. The Gaussians in the
      rightmost part of the figure once again represent the relatively
      poorly determined components, mostly behind the edges of the
      observed profiles and therefore this criterion partly duplicates
      the second and the third criterion.

   \item The Gaussians that are located to the right of the green line
      in Fig.~\ref{UFig01}. The measured values of the brightness
      temperatures in the GASS are all less than $160~\mathrm{K}$, but
      some Gaussians are considerably higher than that and must
      therefore correspond to some specific features in the profiles.
      High components, which also have relatively large widths can only
      have their centers outside the velocity range covered by GASS, and
      therefore this criterion duplicates the second and third
      criterion, and also sometimes the fourth. As a rule, very narrow,
      but high Gaussians have their centers between the velocities,
      corresponding to the receiver channels. These components usually
      represent the very steep edges of the absorption features in the
      profiles and therefore cannot be considered as completely
      spurious.

   \item The Gaussians located on the left-hand side of the blue curve
      in Fig.~\ref{UFig01}. These are narrow and weak components with
      small areas under them. Most of these Gaussians are caused by
      the fact that during the decomposition we prefer to decompose the
      strongest random noise peaks into Gaussians rather than losing a
      part of the signal. On average, this choice generates less than
      one small Gaussian per decomposed profile. However, occasionally
      some profiles contain tens of weak and narrow components. In most
      cases, this happens when the character of the residuals of the
      decomposition is considerable different in the baseline and signal
      regions of the profile. Two main causes of these differences are
      the presence of moderate RFI at brightness temperatures
      $T_\mathrm{B} > 0.5~\mathrm{K}$ and the correlator failures.
      Actually, the first signs of the presence of the latter problem
      were discovered just through the application of this selection
      criterion. As these are very small Gaussians, and, in general, the
      number of such Gaussians per profile is more important than the
      area under them, for this criterion, the badness is calculated by
      adding their total number per profile to the usual sum of the
      areas under selected Gaussians.

\end{enumerate}
\renewcommand{\labelenumi}{\arabic{enumi}}

\begin{figure}[!t]
   \centering
   \includegraphics[width=9.cm]{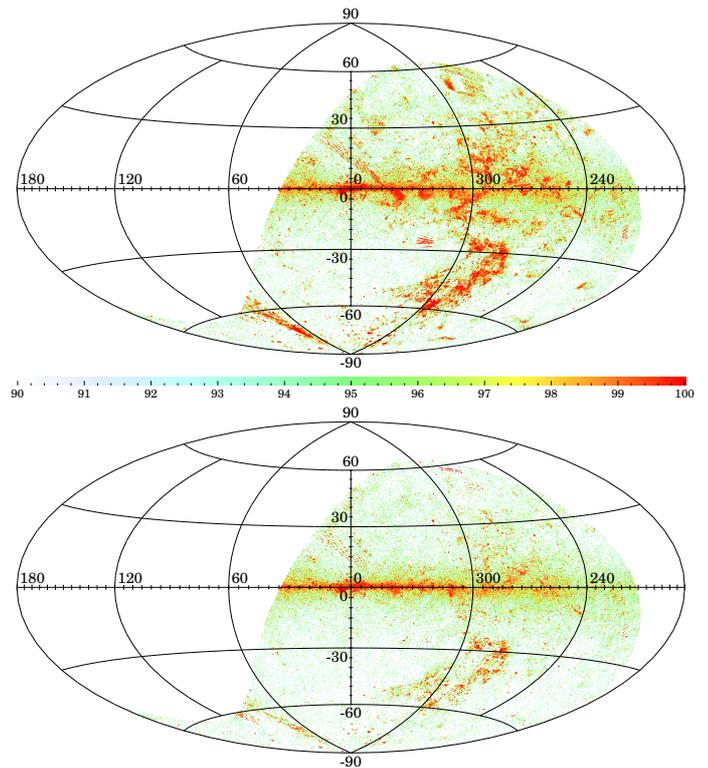}
   \caption{The sky distribution of the most problematic profiles in the
      initial (upper panel) and final (lower panel) database. The color
      scale corresponds to the rank of the badness in the initial data,
      with 100\% corresponding to the most and 0\% to the least
      problematic profile. In the lower panel, each profile is plotted
      with the color that corresponded, in the upper panel, to the value
      of the badness of this profile in the final database.}
   \label{UFig04}
\end{figure}

After using the six criteria described above, and correcting the data on
account of the problems detected by these criteria, we created a short
movie by computing the sky distribution of the brightness temperatures
at running velocities from our latest Gaussian decomposition of the
HEALPix profiles. In this movie, the observational noise is greatly
suppressed and even weak features, approximated by Gaussians, become
clearly visible. In general, the movie appeared as expected, but in the
velocity range of $-350 < V < -250~\mathrm{km\,s}^{-1}$ we found a
number of fields with rather fuzzy borders, which contained brightness
enhancements of the order of $T_\mathrm{B} = 0.075~\mathrm{K}$.

We decided to study these ``stains'' on our movie in greater detail, but
the separation of the Gaussians, responsible for the enhancements, was
not easy. First, we determined that the typical central velocities and
widths of corresponding Gaussians fall into the blue rectangle in
Fig.~\ref{UFig02} ($-350 < V_\mathrm{C} < -250~\mathrm{km\,s}^{-1}$ and
$11.7 < \mathrm{FWHM} < 50~\mathrm{km\,s}^{-1}$), but similar components
may also represent high-velocity clouds (HVCs; the horizontal band of
enhanced frequency of the Gaussians at about $\lg(\mathrm{FWHM}) = 1.4$)
and there is also an enhancement of narrower components at the same
velocities down to the very narrow Gaussians, mostly representing the
higher peaks of the random noise (orange bands at about
$\lg(\mathrm{FWHM}) = -0.15$ and $0.1$).

To exclude the noise from the following discussion, we use the cluster
analysis, as described in Sec. 2 of \citet{Hau10}, and consider in the
following only the clusters of at least three Gaussians in which the
average velocities and $\mathrm{FWHM}$ fall inside the limits, indicated
in Figs.~\ref{UFig02} and \ref{UFig03} by blue lines. In doing so, we
assume it to be rather improbable that three neighboring profiles have
similar random noise peaks. We also reject all small clusters at great
distances from the large clusters at the centers of the stains. To
exclude the HVCs, we use the fact that at considered velocities the HVCs
are mostly located in the longitude range $5\degr < l < 130\degr$ and
the stains were outside this region. However, to be on the safe side, we
also do not search for stains at up to $15\degr$ around the South
Galactic pole, $20\degr$ around the Galactic center, and $3\degr$ around
the Galactic plane.

As a result, we obtained a list of 1172 clusters, containing 24867
Gaussians, which we could use for the search of problematic data dumps.
It turned out that observations between January and June 2005 are
affected by low-level RFI. This period is part of the first coverage
with declination scans. The corresponding data from the second coverage,
scanning in RA, were found to be unaffected by this kind of RFI. We
flagged all channels for dumps from the first coverage, which
contributed to positions in the list of problematic Gaussians. The RFI
usually affects most of the multibeam systems at the same time; for this
reason the flagging was applied simultaneously to all receivers whenever
valid data from the second coverage were available. We corrected 201929
individual dumps by flagging 14720316 affected channels in total.

\begin{figure}[!t]
   \centering
   \includegraphics[width=9.cm]{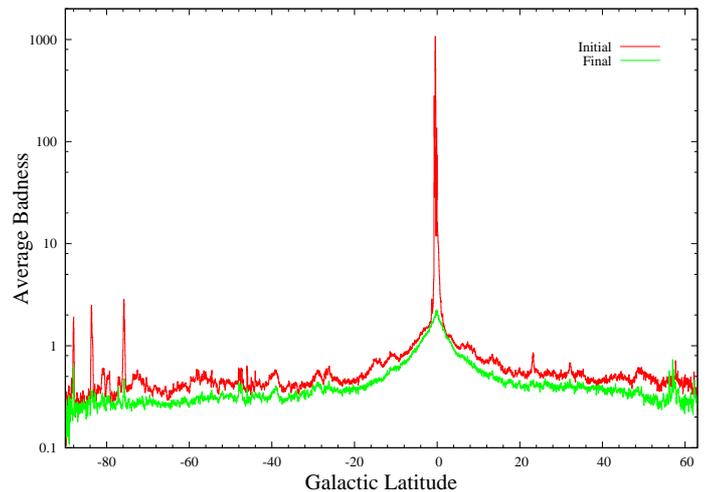}
   \caption{The average badness of the profiles at different Galactic
      latitudes in the initial and final data.}
   \label{UFig05}
\end{figure}

\begin{figure}[!t]
   \centering
   \includegraphics[width=9.cm]{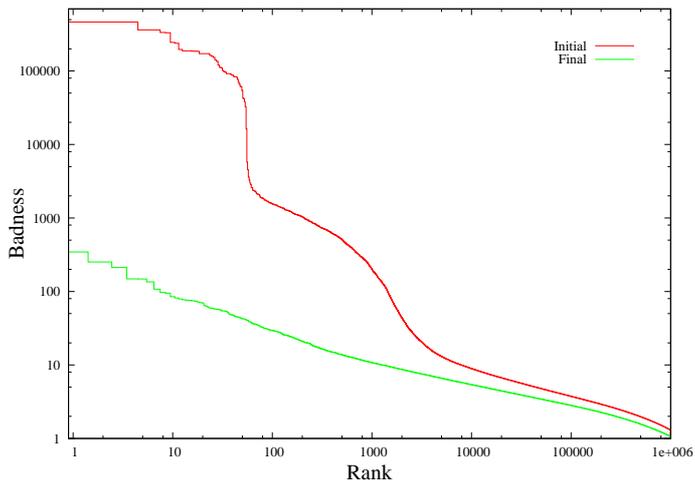}
   \caption{The comparison of the badness of the most problematic
      profiles in the initial and final database. The rank of the
      profile badness in the corresponding database is used as an
      abscissa. On this axis, the value of 640\,126 corresponds to 90\%
      on the color scale of Fig.~\ref{UFig04} and 1 corresponds to 100\%.}
   \label{UFig06}
\end{figure}

As the cluster analysis is a rather labor consuming endeavor, it was
applied only to the version of the data obtained after using the first
six criteria. For the calculations of the badness used in
Figs.~\ref{UFig04} -- \ref{UFig06}, the selections of velocities, line
widths, and sky coordinates for the stains, are applied to individual
Gaussians in the decompositions of the initial and final data sets. This
adds some contamination by random noise to these figures. Nevertheless,
examples of the described stains are well visible in the upper panel of
Fig.~\ref{UFig04} around $(l,~b) = (291\degr,~51\degr)$,
$(229\degr,~24\degr)$, $(263\degr,~4\degr)$, $(207\degr,~-40\degr)$ and
$(198\degr,~-57\degr)$. In the upper panel of Fig.~\ref{UFig02} the
vertical enhancement in the distribution of Gaussians, similar to that
discussed above, is visible also around $V_\mathrm{C} =
234~\mathrm{km\,s}^{-1}$. The weaker enhancements exist near
$V_\mathrm{C} = -420$, $423$, $-476$, and $470~\mathrm{km\,s}^{-1}$. We
checked corresponding movie frames, but as no obvious stains were found,
we did not attempt to flag the corresponding regions of the data dumps.

The results of the searches of the spurious features in the initial and
final profiles are presented in Figs.~\ref{UFig04} -- \ref{UFig06}. The
upper panel of Fig.~\ref{UFig04} illustrates the sky distribution of the
10\% of the profiles with the highest badness values in the initial
database. Here the colors of the points represent the ranks of the
corresponding badness values. The lower panel is for the final data and
the final badness of each profile is given with the same color as was
used for the corresponding value in the upper panel. Therefore, if the
color of some profile in the lower panel is blue shifted with respect to
the color of the same profile in the upper panel, according to our
criteria this profile in the final data is improved in comparison with
its initial version. Shifts toward red indicate that the corresponding
profiles have become even more problematic. Figures \ref{UFig04} and
\ref{UFig05} show that the profiles in the Galactic plane have the
lowest quality (highest badness). There are two reasons for this: the
system noise is highest in this region, and the baseline is most
uncertain because the fit there has to span the largest velocity
intervals. To ensure constant quality at all latitudes, it would have
been necessary to increase the integration time at low latitudes during
observations.

The badness values for the most problematic profiles are given in
Fig.~\ref{UFig06}. We can see that about 5\,000 profiles were
considerably improved (according to our criteria) and smaller
improvements were obtained in many more cases. The bumps in the red
curve for the initial badness distribution can be identified with
residual instrumental problems in GASS II. The improved baseline
algorithm and the rejection of the dumps with the worst correlator
failures has led to a general decrease in the badness. For GASS
III, the slope of the green log-log relation in Fig.~\ref{UFig06} is
approximately constant. Since significant wiggles in the green line are
missing, we may conclude that the most severe problems have probably
been removed.

\begin{figure}[!t]
   \centering
   \includegraphics[angle=-90,width=9.cm]{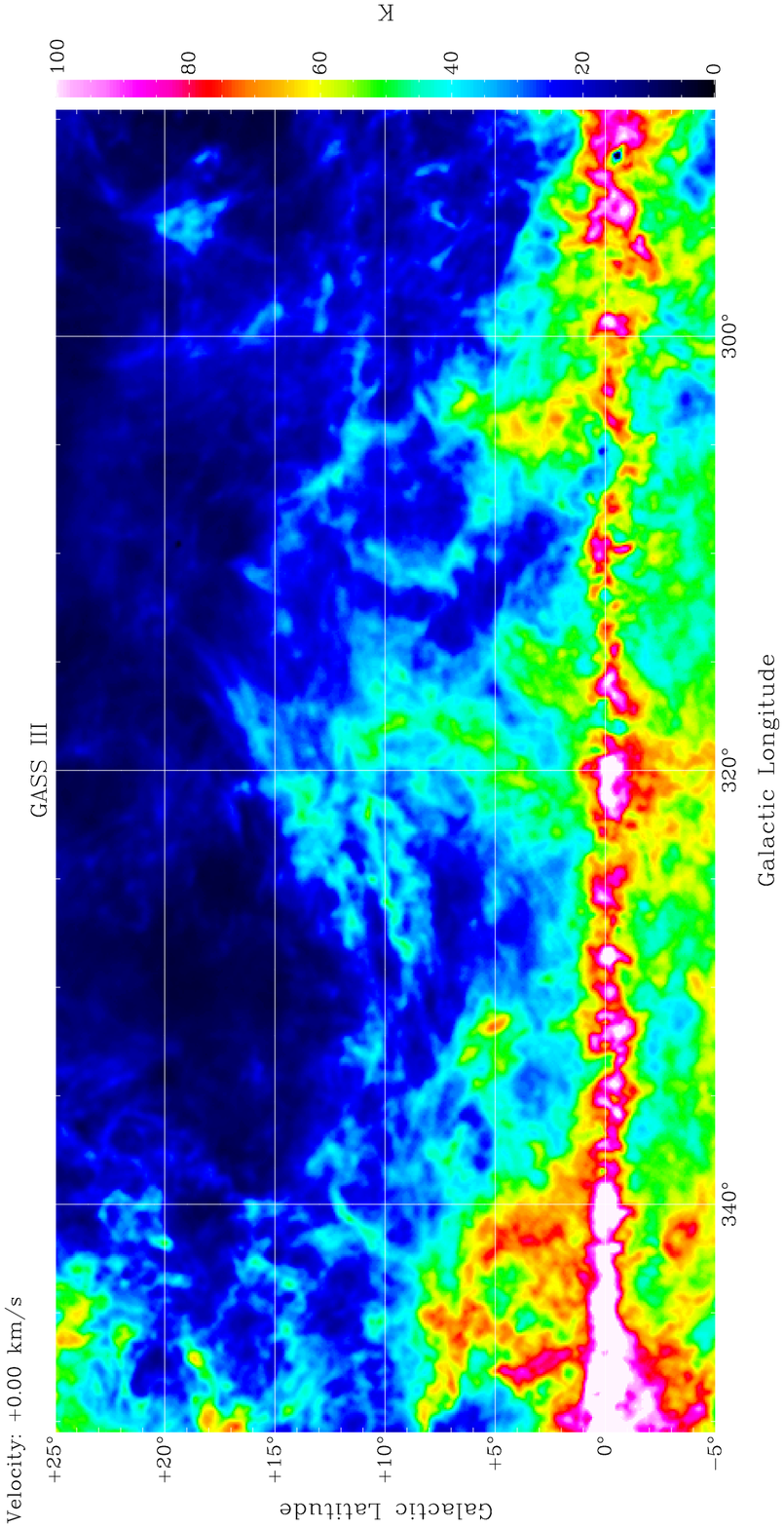}
   \includegraphics[angle=-90,width=9.cm]{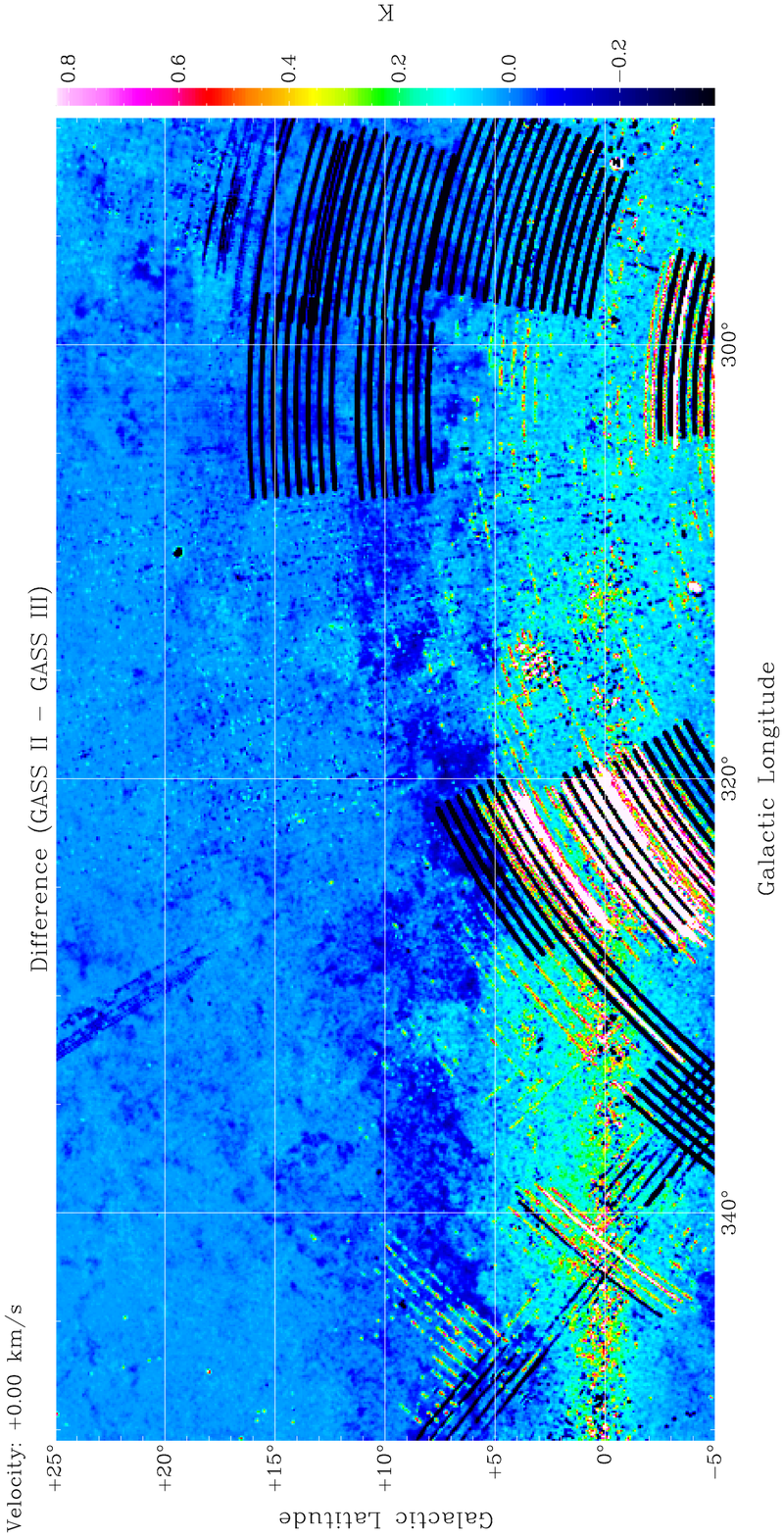}
   \includegraphics[angle=-90,width=9.cm]{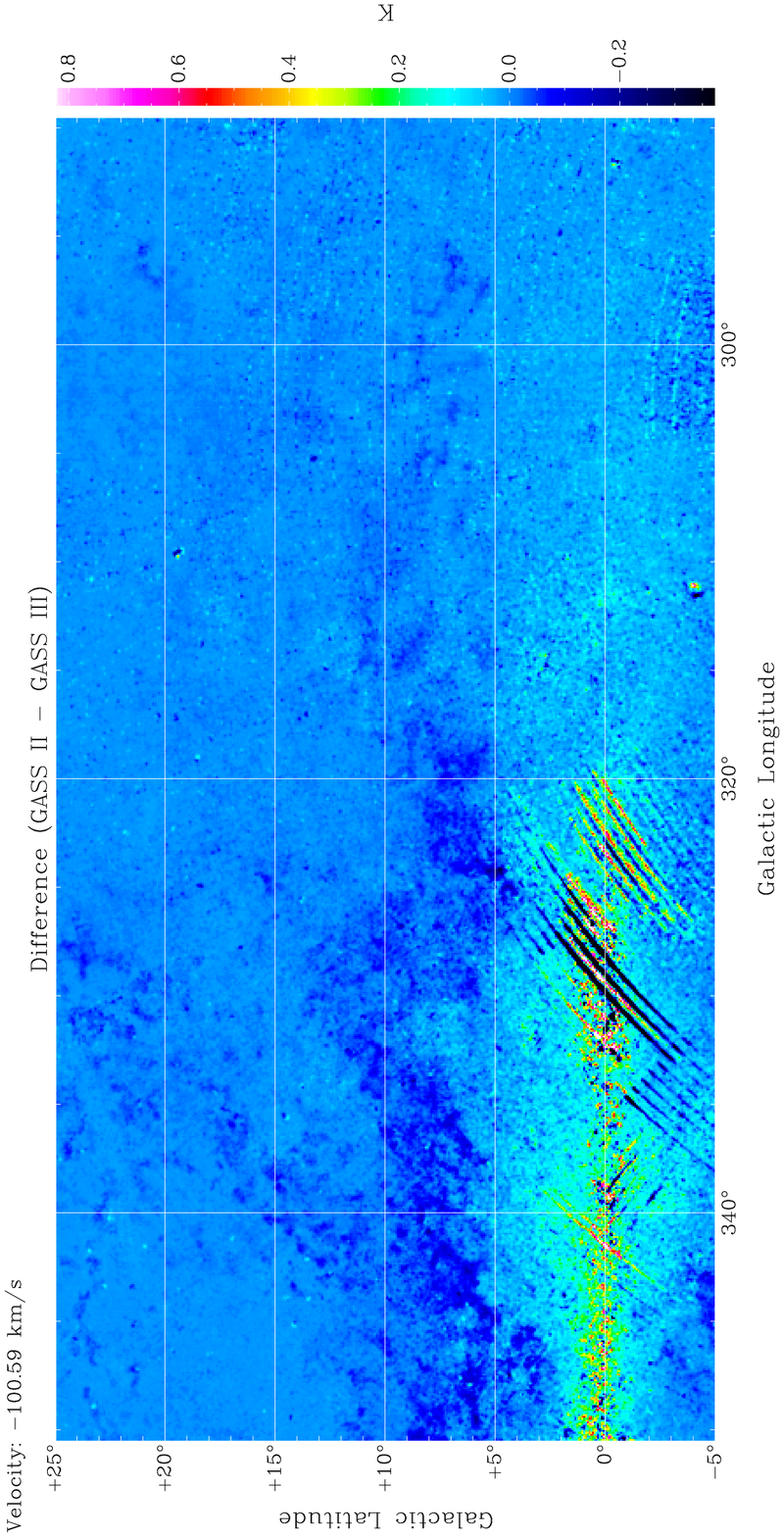}
   \caption{Top: the GASS III HI emission at ${\rm v_\mathrm{lsr}} = 0.0
      $ \kms. For comparison, we show errors that have been removed from
      GASS III; in the middle at the same velocity, at the bottom for $
      v_\mathrm{lsr} = -100.6 $ \kms. This comparison demonstrates that
      most of the instrumental effects are strongly frequency dependent.}
   \label{Fig_RFI_A}
\end{figure}

\subsection{Visualization}
\label{Maps}

Parallel to checking the results of the Gaussian analysis, we calculated
3-D FITS cubes for visual inspection. To demonstrate the progress in
cleaning the GASS database, we present a few channel maps.

Figure \ref{Fig_RFI_A} shows a region that is typical for strong \hi~
emission in the Galactic plane. The GASS III image for $v_\mathrm{lsr} =
0.0 $ \kms~ is displayed on top. Below we show changes in comparison to
GASS II. Most obvious are scanning stripes in the right ascension or
declination direction that originate from RFI or correlator failures. At
$l \sim 315\degr, b \sim 3\degr$ a typical RFI footprint is visible,
most of the other isolated spots are due to RFI. The RFI footprints tend
to shift with velocity along the scanning direction while the stripes
show oscillating intensities. Close to the Galactic plane we find an
average positive baseline offset of $\delta T_B \sim +100$ mK which is
surrounded by a region with an offset of $\delta T_B \sim -50$ mK.
Offsets of this kind are slowly variable and caused by imperfect
baseline fitting in GASS II as discussed in Sect. \ref{Baseline}. The
bottom of Fig. \ref{Fig_RFI_A} at ${v_\mathrm{lsr}} = -100.0 $ \kms
shows that the striped structures are less prominent at high velocities
but baseline offsets are also present there. In total, 2.2\% of the
dumps within this data cube have been discarded. For the whole survey
about 1.5\% of the dumps are affected.

\begin{figure}[!ht]
   \centering
   \includegraphics[angle=-90,width=9.cm]{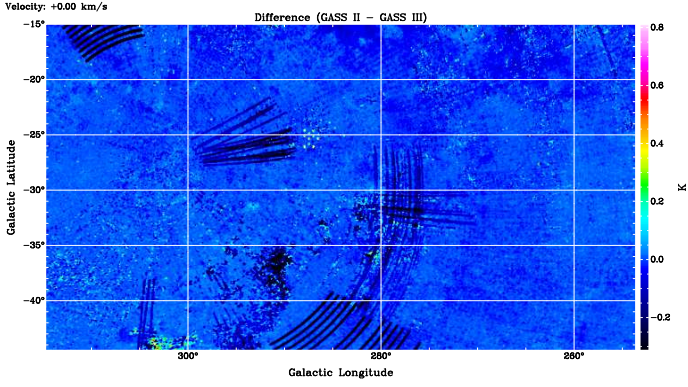}
   \caption{Instrumental errors, eliminated from GASS III, at
      $v_\mathrm{lsr} = 0.0 $ \kms, typical for high Galactic latitudes.
      The stripes are in scanning direction, at $l \sim 288\degr, b \sim
      -25\degr$ a typical RFI footprint is visible, some regions show
       baseline offsets.}
   \label{Fig_RFI_B}
\end{figure}

Figure \ref{Fig_RFI_B} displays corrected errors in a region typical for
high latitudes. At $l \sim 288\degr, b \sim -25\degr$ we find an RFI
footprint. Stripes at high latitudes are in general less prominent due
to low \hi~ brightness temperatures (the correlator problems were found
to scale with line intensity). The south pole is at $ l \sim 303 \degr$,
$ b \sim -27 \degr$, and we demonstrate how the stripes are related to
the telescope scanning in right ascension or declination.

\begin{figure}[!t]
   \centering
   \includegraphics[angle=-90,width=9.cm]{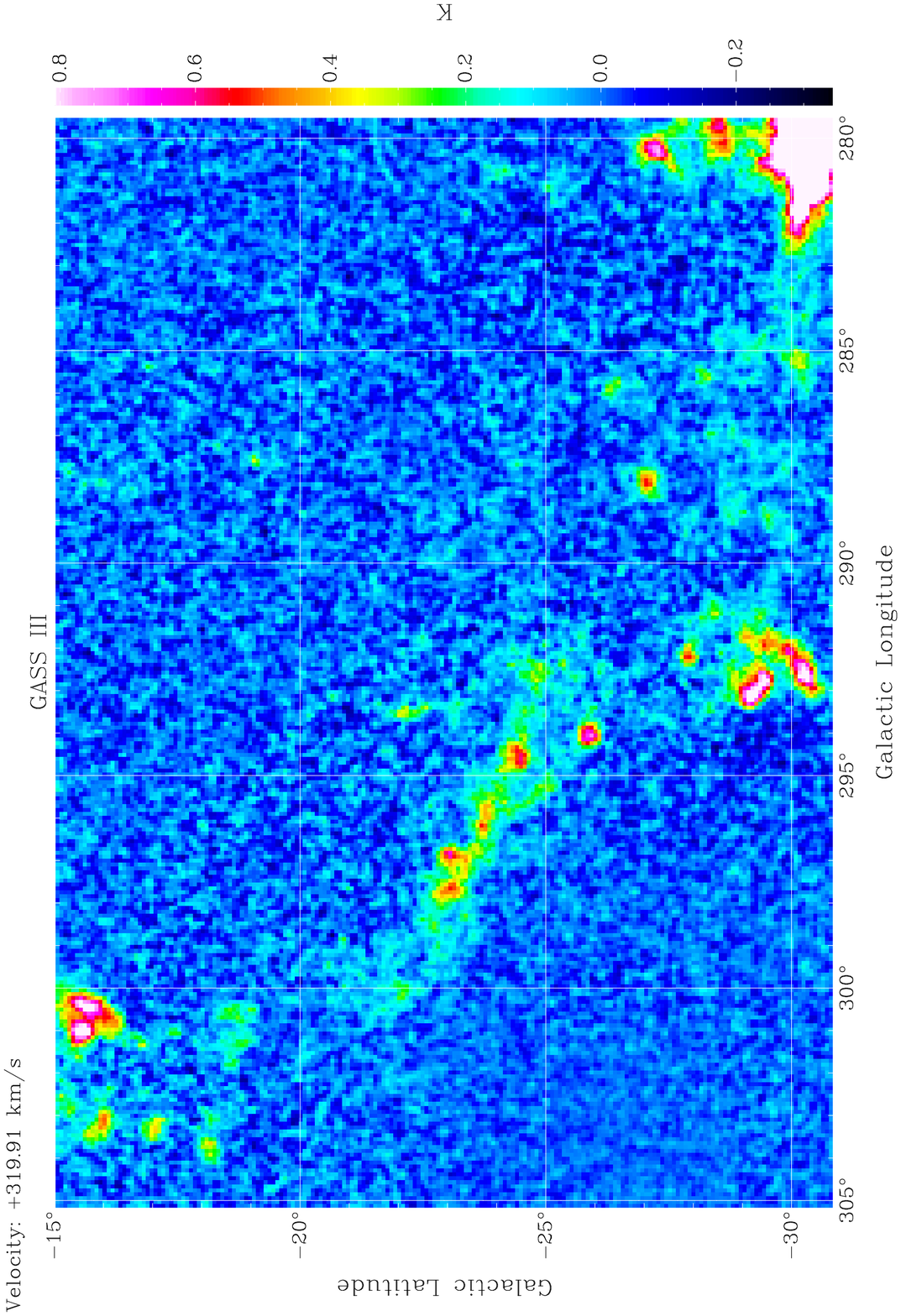}
   \includegraphics[angle=-90,width=9.cm]{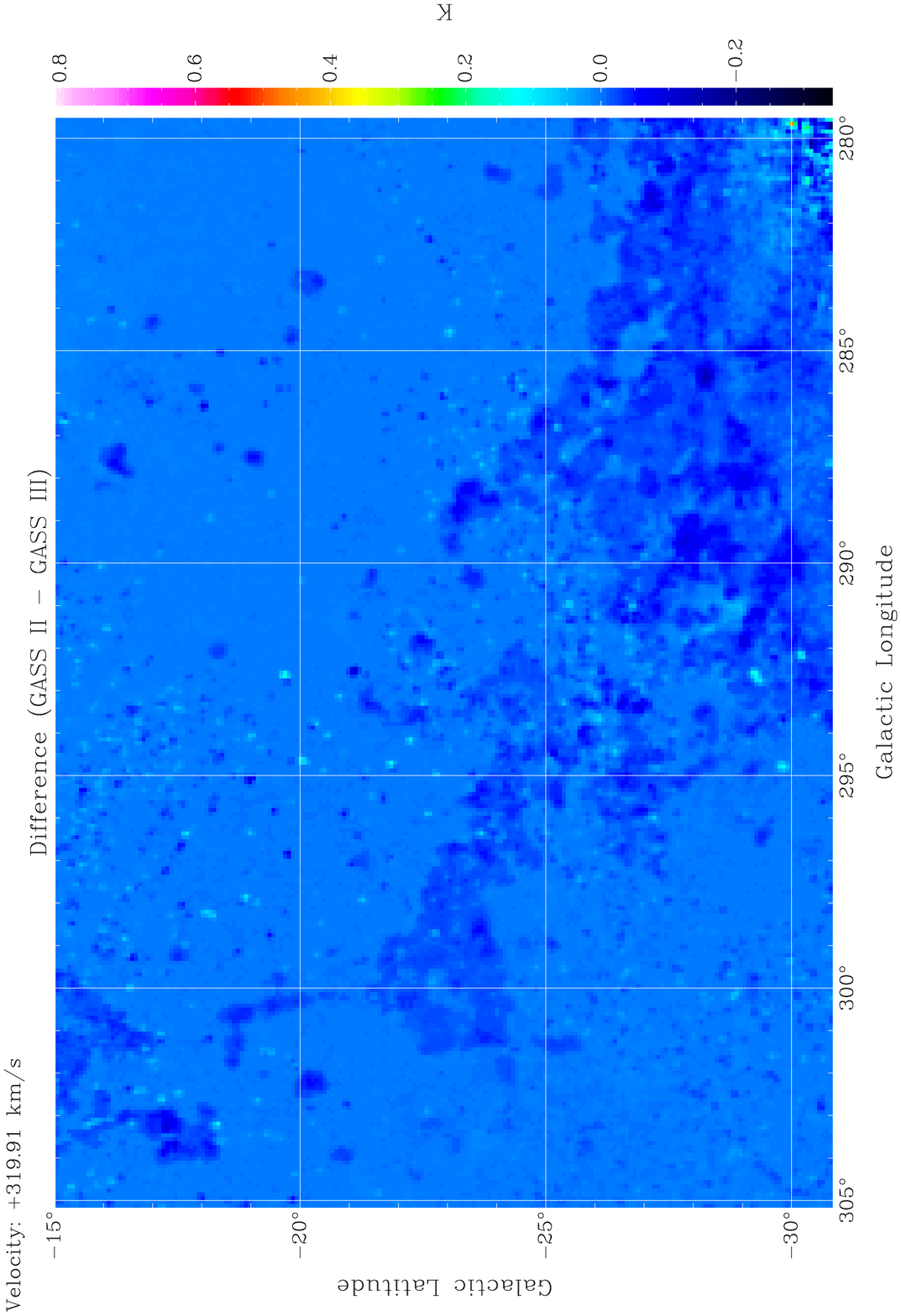}
   \caption{Top: weak emission in the leading arm region. Bottom: RFI at
     individual positions and extended regions with systematical
     baseline offsets that have been removed.}
   \label{Fig_RFI_C}
\end{figure}

Figure \ref{Fig_RFI_C} demonstrates improvements in the baseline. It
shows some low-intensity emission features at ${v_\mathrm{lsr}} = 320.0
$ \kms, belonging to the leading arm of the Magellanic system. Along
these features baseline offsets around -50 mK are visible, which were
caused in GASS II by inappropriate baseline estimates derived from the
LAB survey (see Sect. \ref{Baseline} for details). In GASS III, baseline
biases for emission features of this kind are removed. Weak sources are
better isolated from the background noise, which can lead to a doubling
in area covered by these weak features (Verena L\"ughausen, private
communication).

For a more extended overview about improvements that were obtained with
our current analysis, we calculated FITS cubes for the brightness
temperature distribution of the GASS II and III versions and subtracted
our current $T_B$ results from the previous data release
(\citetalias{Kalberla2010}). We generated a movie that can be
downloaded: https://www.astro.uni-bonn.de/hisurvey/gass/GASS2-3.avi. To
focus on low-level emission, we clipped the data for $-0.05 < T_B < 0.2$
K. The flicker visible for some of the features originates from
channel-to-channel fluctuations that existed for the GASS II data
release.

For a display of the remaining problems, we generated another movie with
the same clip levels for the GASS III data release:
https://www.astro.uni-bonn.de/hisurvey/gass/GASS3noise.avi. This version
emphasizes low-level emission and noise while it saturates for strong
emission features. This way it is possible to observe numerous spurious
low-level features that run across the sky. These features are mostly so
weak that they are hardly visible in individual channel maps. Because of
systematic shifts in velocity they are easily recognizable in a movie.
Pale features mimic emission while dark filaments are due to ghosts. The
velocity shifts are caused by the local standard of rest correction,
which does not apply for terrestrial RFI. All these features are so weak
that they are not recognizable by our cleaning algorithm.

\begin{figure}[!ht]
   \centering
   \includegraphics[width=9.cm]{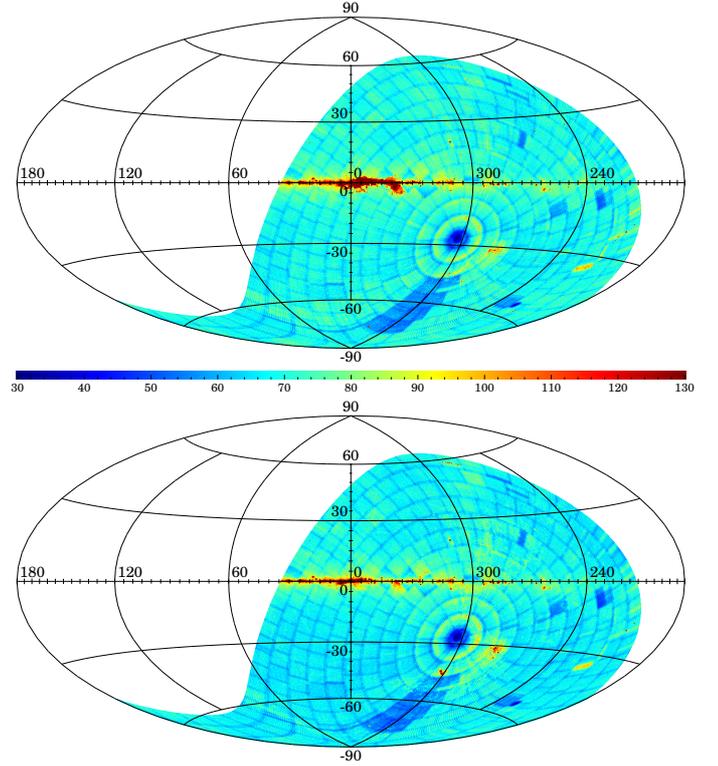}
   \caption{Mean rms noise level in mK across the sky obtained
     from the HEALPix database. Top: GASS II, bottom: GASS III.}
   \label{rms}
\end{figure}

Figure \ref{rms} displays the mean rms uncertainties within a single
velocity channel as determined from the HEALPix database for the initial
GASS II database (top) in comparison to the final result (bottom).
Overall we find that the elimination of unreliable data leads to a
general decrease in the average rms fluctuations by 2.7\%. We only find
a slight increase for very few places. To understand why the elimination
of flagged data leads to a decrease in the uncertainties, we need to
take into account that flagged data usually deviate significantly from a
random distribution. Elimination of bad data leads to an improved error
distribution, approaching the expected white noise.

Figure \ref{rms} only shows mean rms uncertainties. Depending on
flagging of individual velocity channels according to Eq. \ref{Eq3}, the
noise may be different for those regions or velocities that were
seriously affected by RFI or other instrumental problems. For FITS data
products, the noise depends on the map parameters (see
\citetalias{Kalberla2010}, Sect. 6), and for this reason the noise level
of Fig. 9 in \citetalias{Kalberla2010} differs from that of Fig.
\ref{rms}.

\section{Calibration issues}

\subsection{Comparison between GASS and LAB}

\begin{figure}[!t]
   \centering
   \includegraphics[angle=-90,width=9.cm]{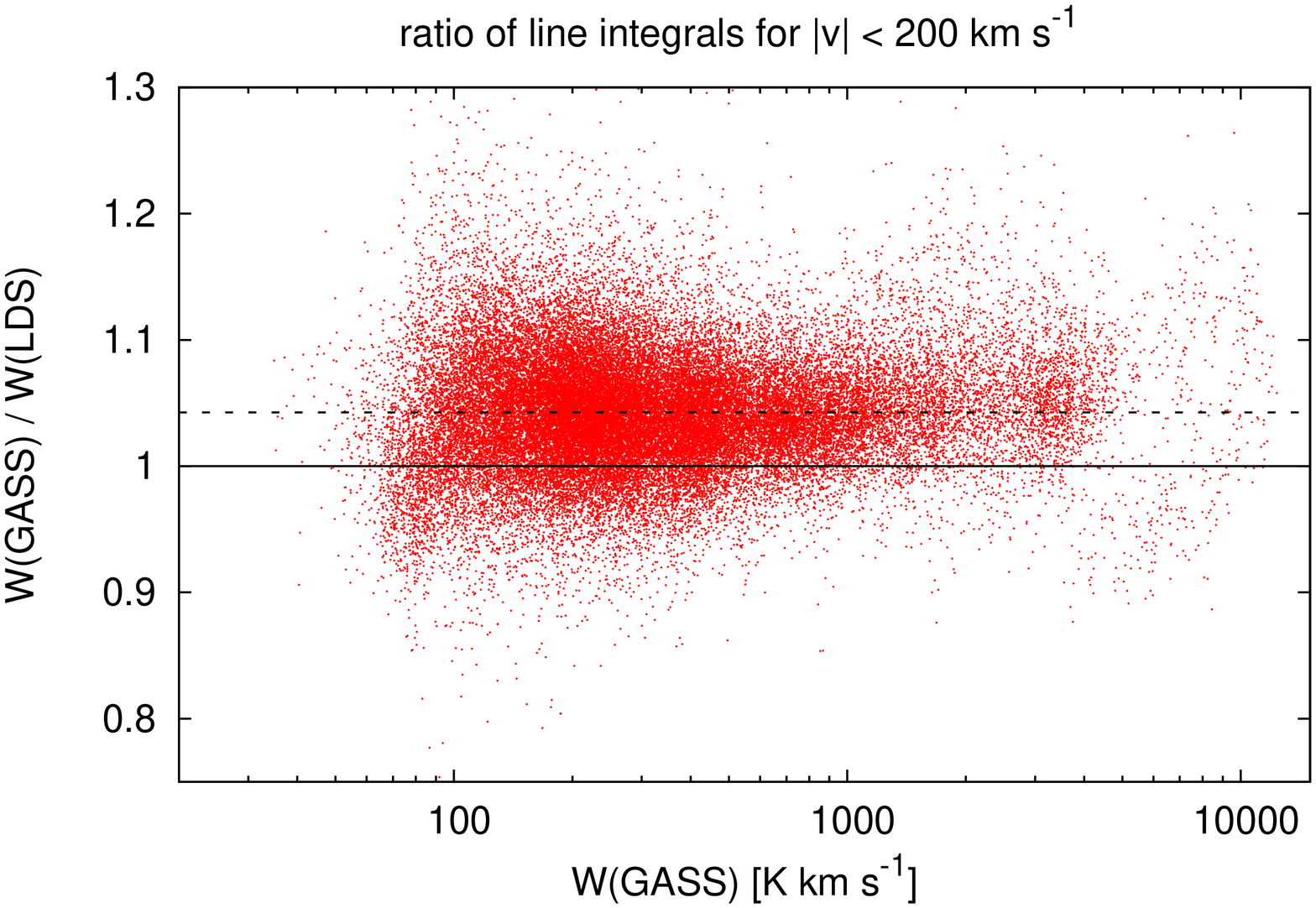}
   \includegraphics[angle=-90,width=9.cm]{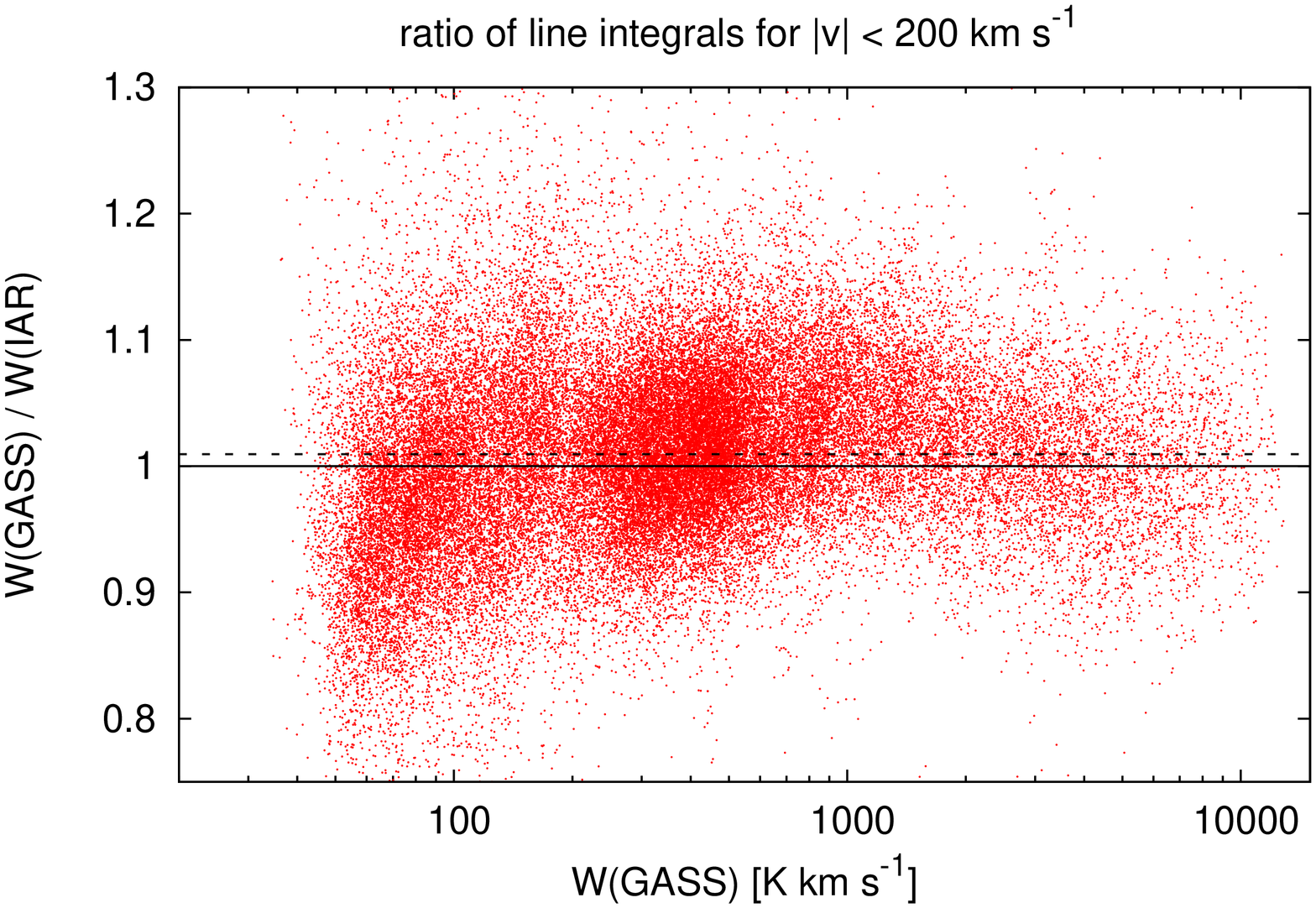}
   \caption{Comparison of line integrals $W_\mathrm{GASS III} /
      W_\mathrm{LAB}$ derived for $|v_\mathrm{lsr}| < 200.0 $ \kms from
      the LAB survey and from GASS III. We distinguish between data
      observed with the 25-m Dingeloo telescope (LDS, top) and with the
      30-m IAR telescope at Villa Elisa (IAR, bottom). The dashed lines
      are for $W_\mathrm{GASS III} / W_\mathrm{LDS} = 1.0426 $ (top) and
     $W_\mathrm{GASS III} / W_\mathrm{IAR} = 1.0096 $ (bottom).}
   \label{Fig_W}
\end{figure}

To compare the complete GASS III with other surveys for quality control,
there is essentially only the LAB survey available. We therefore
generated for both surveys a database of average profiles on a HEALPix
grid with $N_\mathrm{side} = 128$. These profiles were then integrated
for $|v| < 200 $ \kms for comparison.

The calibration of the GASS was independent from the LAB (see Sect. 4 of
\citetalias{Kalberla2010}), however, baselines of GASS II were derived
by bootstrapping from the LAB survey, resulting in possible biases. The
GASS III has improved baselines that were derived iteratively without
using LAB data. Therefore LAB and GASS III are independent from each
other.

The LAB survey was observed with two different telescopes, and
accordingly, we compare both data sets separately. Declinations $ \delta
> -27\fdg5$ were observed with the 25-m Dingeloo radio telescope
\citep[Leiden-Dwingeloo-Survey, LDS, ][]{Atlas1997}, while declinations
$ \delta \le -27\fdg5$ were observed with the 30-m telescope of the
Instituto Argentino de Radioastronom{\'i}a (IAR) by \citet{Bajaja2005}.

\begin{figure}[!ht]
   \centering
   \includegraphics[width=9.cm]{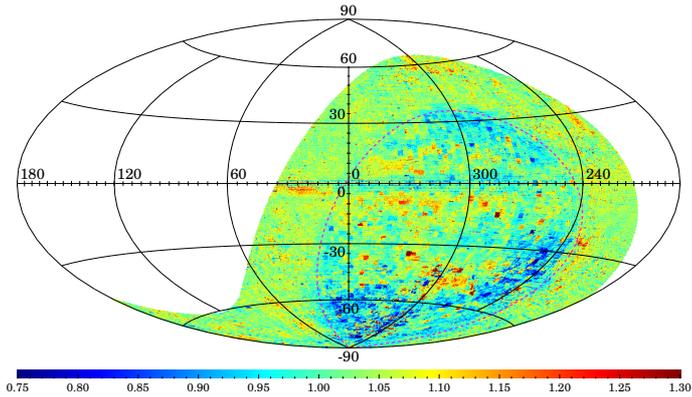}
   \caption{Spatial distribution of $W_\mathrm{GASS III} /
      W_\mathrm{LAB}$. Declinations $ \delta \ge -27\fdg5$ (indicated by
      the dashed line) were observed with the Dwingeloo telescope and
      have systematically high $W_\mathrm{GASS III} / W_\mathrm{LAB}$
      values.}
   \label{Fig_Wdist}
\end{figure}

Figure \ref{Fig_W} displays the ratios of the line integrals
$W_\mathrm{GASS III} / W_\mathrm{LDS}$ (top) and $W_\mathrm{GASS III} /
W_\mathrm{IAR}$ (bottom) as a function of the line integrals
$W_\mathrm{GASS III}$. Fitting average quotients, we obtain
$W_\mathrm{GASS III} / W_\mathrm{LDS} = 1.0426 (\pm 0.0002)$ and
$W_\mathrm{GASS III} / W_\mathrm{IAR} = 1.0096 (\pm 0.0004)$. For the
whole LAB survey, we obtain $W_\mathrm{GASS III} / W_\mathrm{LAB} =
1.0254 (\pm 0.0002)$. The scatter visible in Fig. \ref{Fig_W} is
considerable, but because of the large number of data points ($n=48824$
and $n=52912$ for LDS and IAR, respectively) the appearance is somewhat
biased by outliers. The quotients are well defined and show systematic
differences between LDS and IAR.

To disclose whether there might be a systematic trend in the spatial
distribution of these inconsistencies we plot in Fig. \ref{Fig_Wdist}
$W_\mathrm{GASS III} / W_\mathrm{LAB}$ in Galactic coordinates.
Different systematic effects for the ratios $W_\mathrm{GASS III} /
W_\mathrm{LAB}$ for the LDS and IAR survey are obvious.

For declinations $ \delta > -27\fdg5$, the region covered by the LDS
survey, there appear to be fluctuations without a clear trend. For lower
declinations, covered by the IAR, there is a range with systematically
low values of $W_\mathrm{GASS III} / W_\mathrm{IAR} $ but there are also
apparently no systematic trends. Fluctuations for declinations $ \delta
\le -27\fdg5$ are significantly larger than for declinations above this
limit. We find blocky structures in Fig. \ref{Fig_Wdist}, and in Fig. 8
at the bottom of \citetalias{Kalberla2010}, indicating that the errors
are correlated with the observing strategy. For the IAR, a grid of 5 by
5 positions was observed. The blocky structure is found to be frequency
dependent, the IAR data may therefore contain some residual stray
radiation or baseline problems.

\subsection{Comparison with other telescopes}
\label{Cal}

The GASS brightness temperature scale may also be compared with
calibrations at IAU standard position S8 \citep{Williams1973} as
determined by \citet{Kalberla1982} for the Effelsberg telescope. For the
S8 line integral within the velocity range $ -5.1 < v_\mathrm{lsr} <
22.3 $ \kms~ , we obtain $W_\mathrm{GASS III} / W_\mathrm{Eff} = 1.033
\pm 0.005$. Comparing the peak temperatures, we get $T_\mathrm{GASS III}
/ T_\mathrm{Eff} = 1.041 \pm 0.01$. We conclude that the GASS III
brightness temperature scale is about 4\% higher in comparison to the
scale determined by the Dwingeloo and Effelsberg telescopes in the
northern hemisphere.

The Dwingeloo brightness temperature scale dates back to the calibration
by \citet{Baars1977} against Tau A and is within uncertainties of $< 1\%
$ consistent with the LDS scale \citep{Kalberla2006}. A similarly good
agreement was found with the independently determined Effelsberg scale
\citep{Kalberla1982}.

Apparently, there is a systematic mismatch between the calibrations in
the northern and southern hemispheres. We use results from an
independent absolute calibration obtained with the Green Bank Telescope
(GBT) with the hopes of shedding light on the origin of this
discrepancy. \citet[][Sects. 5.2, 7.4, 8.1, and 8.2]{Boothroyd2011}
compare their calibration with Effelsberg and LAB data. They find for S8
$W_\mathrm{Eff} / W_\mathrm{GBT} = 1.018 \pm 0.018$, $W_\mathrm{LAB} /
W_\mathrm{GBT} = 1.019 \pm 0.007$, and for S6 $W_\mathrm{LAB} /
W_\mathrm{GBT} = 1.015 \pm 0.010$. In addition, a number of pointed
observations yield $W_\mathrm{LAB} / W_\mathrm{GBT} = 1.0288 \pm
0.0012$. Thus the LAB and Effelsberg brightness temperature scales are
consistently $2 - 3$\% higher in comparison with the GBT. This implies,
however, that the GASS brightness temperature scale exceeds the GBT
scale by 6-7\%.

After this more extended comparison of differences in the calibration
between several telescopes, we arrive at the conclusion that there must
be a systematic bias in calibration between the northern and southern
hemisphere. The discrepancy most probably reflects calibration
uncertainties that remained undetected previously since observers in
both hemispheres tend to use different standard positions for their
internal calibration: S7 in the north and S9 in the south. Systematic
calibration errors are, in any case, hard to estimate. We need to accept
uncertainties of 1 -- 3\%.

On one hand, we have in the north the Dwingeloo, Effelsberg, and GBT
telescopes, which have within one or two percent a common consistent
brightness temperature scale. In the south, the calibration between the
IAR and the Parkes telescopes is also very consistent within one
percent. The problem is that both scales differ systematically by 3 --
4\%. This discrepancy exceeds the expected uncertainties. To obtain a
consistent temperature scale, we propose to match the GASS brightness
temperatures to the LAB survey by scaling down the GASS brightness
temperatures by a factor of 0.96. This scale would also be consistent
with the Effelsberg calibration \citep{Kalberla1982} and the remaining
deviations from the GBT scale would be limited to $\sim 2$\%.

\section{Summary}

The second edition of the GASS has been in use since 2010 and it has
become evident that some instrumental problems remained to be solved. We
developed an iterative procedure to remove these problems as far as
possible.

In the first instance, we generate a database of profiles averaged on a
HEALPix grid with $ N_\mathrm{side} = 1024$. For quality control these
profiles are decomposed into Gaussian components. When fitting
instrumental baselines of individual telescope dumps we use priors from
the HEALPix database. Next we calculate whether the rms deviations from
the nearest average HEALPix profile are acceptable for each dump. We
search the dumps for oscillations caused by correlator failures or RFI.
Bad dumps are marked and excluded from calculating averages in the
further iterations. Eventually, we generate 3-D FITS cubes for visual
inspection and proceed to calculate a new version of improved averages
for the next iteration. In total, we need to exclude about 1.5\% of the
telescope dumps for an improved version of the GASS. Examples for the
improvements are given in Figs. \ref{Fig_RFI_A} to \ref{Fig_RFI_C}.
Essential for this progress was the {\it \textup{combined}} analysis of
Gaussian components and the visual inspection of 3-D FITS data cubes for
the complete survey.

We compare GASS III with the LAB survey and find differences that most
probably indicate systematical deviations in calibration. For
declinations $ \delta > -27\fdg5$, the GASS III brightness temperature
is 4\% too high in comparison with the Dwingeloo and Effelsberg scales,
south of this region the GASS III scale is 1\% too high in comparison
with the IAR survey. In comparison to the GBT calibration we even find a
discrepancy of 6-7\%.

Data products from GASS III, compatible with the GASS II web-interface,
are available at http://www.astro.uni-bonn.de/hisurvey/.


   \begin{acknowledgements}
      We acknowledge the referee's detailed and constructive criticism.
      PK acknowledges Jürgen Kerp and Benjamin Winkel for support and
      technical discussions, and Benjamin Winkel for pointing out
      discrepancies in the calibration of the telescopes discussed in
      Sect. \ref{Cal}. PK also acknowledges continuous support at AIfA
      after retirement, including generous allocation of computing
      resources. UH was supported by institutional research funding
      IUT26-2 of the Estonian Ministry of Education and Research and by
      Estonian Center of Excellence TK120. The Parkes Radio Telescope is
      part of the Australia Telescope, which is funded by the
      Commonwealth of Australia for operation as a National Facility
      managed by CSIRO.
   \end{acknowledgements}

\end{document}